\documentclass[a4paper,11pt,twoside]{article}
\pdfoutput=1

\usepackage{geometry} 
\usepackage{a4wide}

\usepackage{amssymb,amsmath,bm}
\usepackage{cite}
\usepackage{color}
\usepackage{slashed}
\usepackage{graphicx}
\usepackage[latin1]{inputenc}
\usepackage{amsfonts}
\usepackage{lscape}
\usepackage{amsthm}
\usepackage{booktabs}
\usepackage{array}
\usepackage{rotating}
\usepackage[numbers, sort]{natbib}
\usepackage{float}
\usepackage{xspace}

\usepackage{mathrsfs}

\usepackage[small,bf]{caption}
\selectfont

\newcommand{\nt}{\widetilde{\chi}^0}
\newcommand{\ch}{\widetilde{\chi}^\pm}
\newcommand{\chp}{\widetilde{\chi}^+}
\newcommand{\chm}{\widetilde{\chi}^-}

\newcommand{\prospino}[0]{\texttt{Prospino~2}\xspace}
\newcommand{\susyhit}[0]{\texttt{SUSY-HIT}\xspace}

\newcommand{\herwig}[0]{\texttt{Herwig++}\xspace}

\newcommand{\delphes}[0]{\texttt{Delphes 3}\xspace}
\newcommand{\checkmate}[0]{\texttt{CheckMate}\xspace}

\newcommand{\BR}{{\rm BR}}

\def\tev{\ensuremath{~\text{TeV}}}
\def\gev{\ensuremath{~\text{GeV}}}

\def\missingET{\ensuremath{\displaystyle{\not}E_T}}

\begin{document}

\begin{titlepage}

\vspace*{-15mm}
\begin{flushright}
ULB-TH/14-15,
LPN14-119,
ZU-TH35/14,
STUPP-14-220
\end{flushright}
\vspace*{0.7cm}

\begin{center}

{
\bf\LARGE
LHC Tests of Light Neutralino Dark Matter without Light Sfermions
}
\\[8mm]
Lorenzo~Calibbi $^{\star}$
\footnote{E-mail: \texttt{lcalibbi@ulb.ac.be}},
Jonas~M.~Lindert $^{\dagger}$
\footnote{E-mail: \texttt{lindert@physik.uzh.ch}},
Toshihiko~Ota $^{\ddag}$
\footnote{E-mail: \texttt{toshi@mppmu.mpg.de}},
Yasutaka~Takanishi $^{*}$
\footnote{E-mail: \texttt{yasutaka@mpi-hd.mpg.de}}
\\[1mm]
\end{center}
\vspace*{0.50cm}
\centerline{$^{\star}$ \it
Service de Physique Th\'eorique, Universit\'e Libre de Bruxelles,}
\centerline{\it
Bld du Triomphe, CP225, B-1050 Brussels, Belgium}
\vspace*{0.2cm}
\centerline{$^{\dagger}$ \it
Physik-Institut, Universit\"at Z\"urich,}
\centerline{\it
Wintherturerstrasse 190, CH-8057 Z\"urich, Switzerland}
\vspace*{0.2cm}
\centerline{$^{\ddag}$ \it
Department of Physics, Saitama University,}
\centerline{\it
Shimo-Okubo 255, 338-8570 Saitama-Sakura, Japan}
\vspace*{0.2cm}
\centerline{$^*$ \it
Max-Planck-Institut f\"ur Kernphysik,}
\centerline{\it
Saupfercheckweg 1, D-69117 Heidelberg, Germany}

\vspace*{1.20cm}
\begin{abstract}
We address the question how light the lightest MSSM neutralino 
can be as dark matter candidate in a scenario where all 
supersymmetric scalar particles are heavy.
The hypothesis that the neutralino accounts for the observed dark matter
density sets strong requirements on the supersymmetric spectrum, thus
providing an handle for collider tests. In particular for a lightest
neutralino below $100$ GeV the relic density constraint translates 
into an upper bound on the Higgsino mass parameter $\mu$ in case all 
supersymmetric scalar particles are heavy.
One can define a simplified model that highlights only the necessary features of
the spectrum and their observable consequences at the LHC. Reinterpreting
recent searches at the LHC we derive limits on the mass of the 
lightest neutralino that, in many cases, prove to be more constraining 
than dark matter experiments themselves.
\end{abstract}

\end{titlepage}


\section{Introduction}
\label{sec:intro}
\addtocounter{footnote}{-4}

Employing the data collected at 7 and 8 TeV of center of mass
energy, the LHC experiments have recently published the results
of an impressive number of searches for electroweak production
of new physics. In many cases, they were able to set constraints
on the masses of new electroweakly-interacting particles above
the previous best bounds from LEP. This is the case in particular for the
electroweak sector of the minimal supersymmetric standard model
(MSSM), as well as of any of its extensions. The exact bounds
depend on the details of the spectrum, especially on the mass
hierarchy controlling the decay chains, and there is a generic
loss of sensitivity in the regime of low mass splittings.
However, it is remarkable that, in the most favourable cases,
the limits in the MSSM are up to 300 GeV for the sleptons
\cite{Aad:2014vma,Khachatryan:2014qwa} and up to 700 GeV for the
charginos and neutralinos
\cite{Aad:2014nua,Khachatryan:2014qwa}.

The above mentioned searches have a crucial role in testing
supersymmetric Dark Matter (DM) scenarios as they allow to probe
the relevant parameter space independently of the colored sector
of the theory, which might in principle be too heavy to be
directly accessed by the LHC experiments. The cardinal idea is
the following: the measurements of the DM relic density based on
Cosmic Microwave Background (CMB) 
observations set non-trivial requirements on the
supersymmetric spectrum, thus providing an handle for collider
tests. This is true in particular if the lightest supersymmetric
particle (LSP) is a bino-like neutralino, whose weak
interactions typically lead to overproduction in the early
universe, unless an efficient annihilation mechanism is at work.
Since a limited set of supersymmetric particles and parameters
is involved in the computation of the neutralino annihilation
cross section, and hence of its relic density, one can define
simplified models that highlight only the necessary features of
the spectrum and their observable consequences at the LHC.

The above sketched procedure has been recently employed by us to
answer the question on how light the MSSM neutralino is still
allowed to be by direct searches for electroweakly-interacting
supersymmetric particles at the LHC
\cite{Calibbi:2013poa,Calibbi:2014coa}. Other related studies on
light neutralino Dark Matter have been recently published in
\cite{Arbey:2013aba,Belanger:2013pna,Hagiwara:2013qya}. For
neutralinos lighter than about 30 GeV, the typical spectrum
selected by the relic density constraints features rather light
staus and Higgsinos, with masses smaller than few hundred GeV
\cite{Belanger:2012jn}. The electroweak production of these
particles and the following decays lead to events with multiple
taus and missing transverse momentum. Employing an ATLAS search
for such a signature in combination with the limits on the decay
rate of the Higgs into neutralinos, we could set a lower bound
on the DM mass at about 24 GeV. Remarkably, with the above
exercise, we showed that electroweak LHC searches are at the
moment more powerful than direct and indirect searches in
testing light neutralino Dark Matter. For early works addressing
limits on light neutralino Dark Matter, see e.g.
\cite{Hooper:2002nq,Belanger:2002nr,Bottino:2002ry}, and for
limits on (very) light neutralinos without cosmological bounds
we refer to Refs. \cite{Dreiner:2009ic,Dreiner:2009er} and
references therein.

In the present paper, we want to extend our previous work to the
case where no light sfermions are in the spectrum,
i.e.~scenarios with only neutralinos and charginos lighter than
few hundreds GeV. A motivation for such an exercise is that
light Higgsinos are the minimal `tree-level' requirement posed
by naturalness arguments. A Higgsino-like LSP can not however
account for the full amount of the observed Dark Matter, unless
its mass is in the TeV range, since the
Higgsino-Higgsino annihilation processes are too efficient,
see e.g.~\cite{Cahill-Rowley:2014boa}. Simultaneous presence of light bino
and Higgsinos is thus the minimal ingredient for electroweak
scale neutralino Dark Matter in natural SUSY. Scenarios with
mixed bino-Higgsino Dark Matter, labelled as `well-tempered
neutralino', can provide a natural DM candidate overcoming the
above mentioned problems of a pure Higgsino (or wino) LSP 
\cite{ArkaniHamed:2006mb}.\footnote{For a recent discussion 
of the LHC prospects of this scenario, see \cite{Bramante:2014dza}.}
We are however interested to focus on the light DM regime
(i.e.~$m_{\nt_1}\lesssim 100$ GeV), where the neutralino can not
be `well-tempered' as it is bounded to be mainly bino due to
chargino mass limits. Let us note in passing that, even giving
up naturalness like in split SUSY scenarios
\cite{Giudice:2004tc,ArkaniHamed:2004fb}, or rather `mini-split'
\cite{Arvanitaki:2012ps} as suggested by the observed Higgs
mass, the set-up we are studying is relevant to obtain the
absolute lower bound on DM mass. In fact, in these models there
are no light sfermions that can mediate the neutralino
annihilation and the relic density requirements must fulfilled
by the gaugino-Higgsino sector alone.

As we are going to see, possible resonant enhancements of the
neutralino annihilation cross section due to s-channel $Z$ and
$h$ exchanges play a crucial role in the low mass regime we are
going to study. This provides a further, purely
phenomenological, motivation for our study: the effective
coupling with nuclei for neutralino Dark Matter close to the
above mentioned resonances might drastically drop, as well as
the today annihilation rate relevant for indirect DM searches,
hence one has to find alternative handles to test this corner of
the parameter space. As we are going to show, if nature has
chosen this peculiar scenario, LHC experiments compete and in
some cases prove to be more constraining than dedicated DM
experiments.

LHC limits and prospects for the gaugino-Higgsino sector of the MSSM have been recently
discussed -- however, without a focus on light neutralino DM -- in
\cite{Bharucha:2013epa,Gori:2013ala,Han:2013kza,Schwaller:2013baa,Han:2014kaa,Han:2014xoa,Martin:2014qra},
including the challenging case of compressed spectra.

The rest of the paper is organized as follows. In section
\ref{sec:resonant} we present the light neutralino parameter
space, where resonant annihilation dominates, including relevant
collider and astrophysical constraints. In section
\ref{sec:pheno} we discuss the resulting LHC phenomenology and
in section \ref{sec:limits} we present the corresponding limits. Finally, in
section \ref{sec:conclusions} we conclude.

\section{Resonant Neutralino annihilations}
\label{sec:resonant}

As anticipated in the introduction, we are interested to study
the phenomenology of the MSSM neutralino as a Dark Matter
candidate in the low-mass regime, i.e.~with $m_{\nt_1}\lesssim
100$ GeV, in the case that only neutralinos and charginos are
possibly light, while the rest of the spectrum, in particular
the sfermions, might be decoupled. This setup is completely
defined by the parameters that describe the gaugino-Higgsino
sector in the MSSM:
\begin{equation}
 M_1,~~M_2,~~\mu,~~\tan\beta,
 \label{eq:parameters}
\end{equation}
which are respectively the SUSY-breaking bino and wino masses,
the superpotential Higgs mixing parameter that controls the
spontaneous electroweak symmetry breaking and sets the mass of
the Higgsinos, and the ratio of the two Higgs doublets vevs.

As a result of the LEP limit on charginos, $m_{\ch_1}\approx
{\rm min}(M_2,~|\mu|) \gtrsim 100$ GeV, the lightest neutralino has
to be mainly bino in the mass range we consider. As usual, an
efficient annihilation mechanism is thus required in order to
satisfy the relic density constraints from CMB observations.
Since we are assuming that there are no sfermions (and no extra Higgs bosons) 
below few
hundreds GeV or more, the main annihilation modes go through
an $s$-channel $Z$ or $h$ exchange:
\begin{equation}
\nt_1\nt_1 ~\to~ Z^*/h^* ~\to ~f \bar{f} 
\end{equation}
Full expressions for the corresponding annihilation cross sections can be found in \cite{Nihei:2002ij}.
Let us recall here that the $s$-wave contribution vanishes in the $h$ mediation case and 
it is suppressed by a factor   $m_{f}^{2}/m_{Z}^{2}$ for a $Z$ exchange. 
On the other hand, $p$-wave contributions are in both cases only suppressed by the temperature, $\sim T/m_{\nt_1}$,
and are therefore relevant for the calculation of annihilation rate in the early universe.

In order to have a qualitative understanding of the dependence of the
relic density on the parameters shown in Eq.~(\ref{eq:parameters}),
we have to consider the interactions of the neutralinos with $Z$ and $h$ only. They are given
by the following expressions \cite{Haber:1984rc}:
\begin{align}
\label{eq:interZ}
 \mathcal{L}_{\widetilde{\chi}^{0}_{i} \widetilde{\chi}^{0}_{j} Z}
&=
 \frac{g}{2c_{W}^{}}
 Z_{\rho}
 \overline{\widetilde{\chi}^{0}_{i}}
 \gamma^{\rho}
 \left[
 O^{Z L}_{ij} {\rm P}_{L}
 +
 O^{Z R}_{ij} {\rm P}_{R}
 \right]
 \widetilde{\chi}^{0}_{j},\\
 \mathcal{L}_{\widetilde{\chi}^{0}_{i} \widetilde{\chi}^{0}_{j} h}
&=
 \frac{g}{2} C^h_{ij}~
 h \overline{\widetilde{\chi}^{0}_{i}} \widetilde{\chi}^{0}_{j},
 \label{eq:interh}
\end{align}
where the couplings are defined as:
\begin{align}
\label{eq:couplZ}
 O^{Z L}_{ij}
&=
-
\frac{1}{2} N_{i3} N_{j3}^{*}
+
\frac{1}{2} N_{i4} N_{j4}^{*},
\quad
O^{Z R}_{ij} = - O^{Z L *}_{ij} \\
 C_{ij}^h
 &=
 \frac{1}{2}
 \left[(N_{i2}-N_{i1}\tan\theta_W)(\sin\alpha N_{j3}+\cos\alpha N_{j4}) + (i\leftrightarrow j)
 \right]. 
 \label{eq:couplh}
\end{align}
The matrix $N$ diagonalizes the neutralino mass matrix:
\begin{equation}
 \nt_i = N_{i1} \widetilde{B} +N_{i2} \widetilde{W}^0  + N_{i3} \widetilde{H}_d^0 + N_{i4} \widetilde{H}_u^0.  
\end{equation}
We refer to the appendix for further details on our conventions and
relevant approximate formulae for the elements $N_{i\alpha}$.

From the expressions in Eqs.~(\ref{eq:interZ}-\ref{eq:couplh}),
we see that the couplings of the lightest neutralino to $Z$ and
$h$ vanish if $\nt_1$ is pure bino (or wino), i.e.~if $N_{13} =
N_{14} = 0$. This only occurs when the Higgsino sector is
decoupled, $\mu \gg M_1,~m_Z$. In fact, using the approximate
expressions shown in the appendix for the Higgsino components 
of $\nt_1$, we find:
\begin{equation}
N_{13} = 
\frac{m_{Z} s_{W}^{}}{\mu}
\left[
 s_{\beta} + c_{\beta} \frac{M_{1}}{\mu}
 \right],
 \quad
 N_{14} =
 -
 \frac{m_{Z} s_{W}^{}}{\mu}
 \left[
 c_{\beta}
 +
 s_{\beta}
 \frac{M_{1}}{\mu}
 \right], 
 \label{eq:Higgsino}
\end{equation}
where for simplicity we assumed $M_2 \gg |\mu|$. Here we defined
$c_\beta \equiv \cos\beta$, $s_\beta \equiv \sin\beta$ and 
$s_W \equiv \sin\theta_W$.
It is therefore
clear that the upper limit on the DM relic density will
translate into an upper limit on $|\mu|$, i.e.~on the mass scale of
the Higgsinos. Thus, relatively light Higgsinos are a generic
prediction of our setup, while from
Eqs.~(\ref{eq:couplZ},~\ref{eq:couplh}) it is clear that the
wino plays no crucial role in the annihilation process
and might in principle be heavier. 

A closer look at the expressions for the annihilation cross section 
reported in \cite{Nihei:2002ij} shows the well-known possibility of a resonant 
enhancement of the $p$-wave annihilation, occurring if
\begin{equation}
 m_{\nt_1}\approx m_Z /2 ~~~{\rm or } ~~~ m_{\nt_1}\approx m_h /2. 
 \label{eq:resonances}
\end{equation}
Obviously, the closer $m_{\nt_1}$ approaches these conditions
the looser the upper bound on $\mu$ becomes, since the
enhancement can compensate smaller couplings of 
$\nt_1\nt_1 Z$, $\nt_1\nt_1 h$. 
On the other hand, we expect the relic density
constraints to set a tighter bound on $\mu$ as the mass $m_{\nt_1}$ lies
further from the resonant conditions of
Eq.~(\ref{eq:resonances}). In what follows, we illustrate and
quantify these simple features by means of a numerical scan of
the relevant parameters (in section~\ref{sec:parameterspace})
and we discuss the LHC phenomenology of this
region of resonant neutralino dark matter (in section~\ref{sec:pheno})
and the constraints set by searches for 
chargino-neutralino production (in section~\ref{sec:limits}).

Before moving on, let us comment about the possible role of the
extra Higgses. It is well known, that an $s$-channel exchange of
the CP-odd Higgs $A$ can also provide an efficient annihilation
mechanism for neutralino DM, especially close to the resonant
condition $m_{\nt_1}\approx m_A/2$ and/or for a sizeable
Higgsino component in $\nt_1$ \cite{Drees:1992am}. However, in the light neutralino
regime we are considering, $A$ would be required to be
relatively light \cite{Bottino:2002ry,Bottino:2004qi,Bottino:2008mf,Fornengo:2010mk}. 
This possibility is challenged \cite{Calibbi:2011ug} by direct
searches for extra Higgses at the LHC \cite{Khachatryan:2014wca}, as well as by the
measurements of the Higgs production and decays, that prove to
be SM-like at least at the $30\%$ level (see e.g.~\cite{Bechtle:2014ewa}), and by rare
decays such as $B_s\to\mu^+\mu^-$ \cite{Aaij:2012nna}. 
For these reasons, here we do not consider
the possibility that the extended Higgs sector of the MSSM plays
a role in the neutralino annihilation and we assume for
simplicity that the heavy Higgses are also decoupled.

\subsection{Constraints and viable parameter space}
\label{sec:parameterspace}

Here we present the results of a random scan of our four
parameters within the following ranges:
\begin{align}
\label{eq:scanparameters}
  20\,\gev \le M_1 \le 80\,\gev,& \quad 100\,\gev \le M_2 \le 1\,\tev, \nonumber  \\
 100\,\gev \le |\mu| \le 1\,\tev,& \quad 5 \le \tan\beta \le 50.
\end{align}
Notice that we scan both signs of $\mu$ while we take $M_i>0$ with no loss of generality: observable effects depend in fact
on the relative sign ${\rm sgn}(\mu M_i)$. 
Furthermore we vary the soft parameters of the stop sector in the following ranges:
\begin{align} 
2~{\rm TeV} < m_{\widetilde t_{L}}, m_{\widetilde t_{R}} \le 5~{\rm TeV} \,\gev ,& \quad -4~{\rm TeV} \le A_t < 4~{\rm TeV}  \,.
\end{align}
Together with the ones in Eq.~(\ref{eq:scanparameters}) these parameters determine the 
value of the physical Higgs mass $m_h$ and thus of the position of the resonance 
in Eq.~(\ref{eq:resonances}).
The other SUSY soft parameters were set to the following constant values:
\begin{align}
\label{eq:fixedparameters}
m_{\widetilde f} = M_3 = m_A = 4~{\rm TeV},~ A_f = 0 \,,
\end{align}
where $m_{\widetilde f}$ represents the remaining sfermion masses, 
$M_3$ is the gluino mass, $m_A$ the
CP-odd Higgs mass, $A_f$ the remaining trilinear couplings. The spectrum has been computed
by means of the routine {\tt SuSpect} \cite{Djouadi:2002ze}, the
branching fractions by the {\tt SUSY-HIT}
package~\cite{Djouadi:2006bz} and {\tt micrOMEGAs}~\cite{Belanger:2013oya,Belanger:2006is,Belanger:2004yn,Belanger:2001fz}
has been used to calculate the neutralino relic density,
as well as the scattering cross section with nuclei and the present thermally-averaged annihilation cross section.\\

\begin{figure}[t]
\centering
 \includegraphics[width=0.48\textwidth]{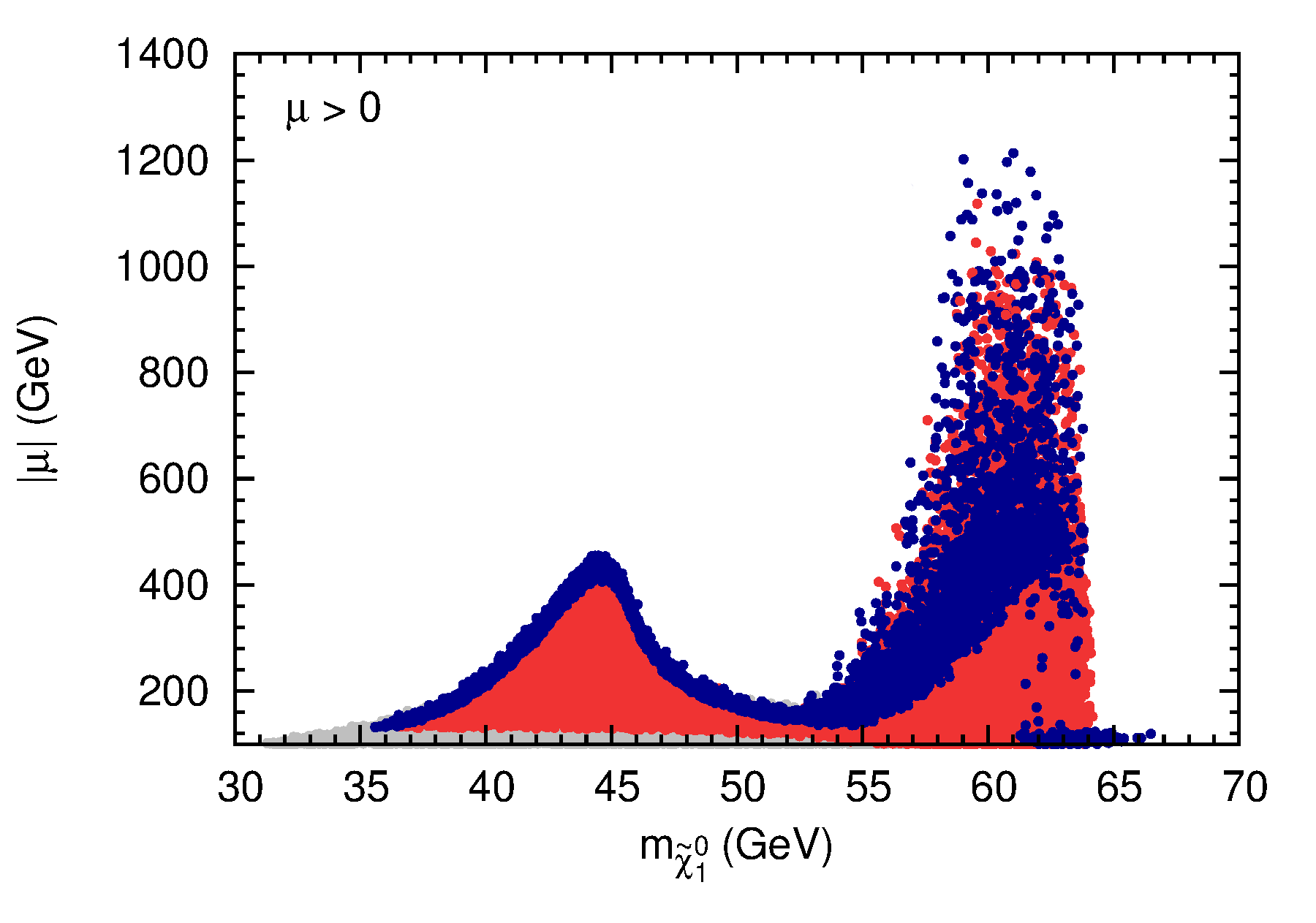}
 \includegraphics[width=0.48\textwidth]{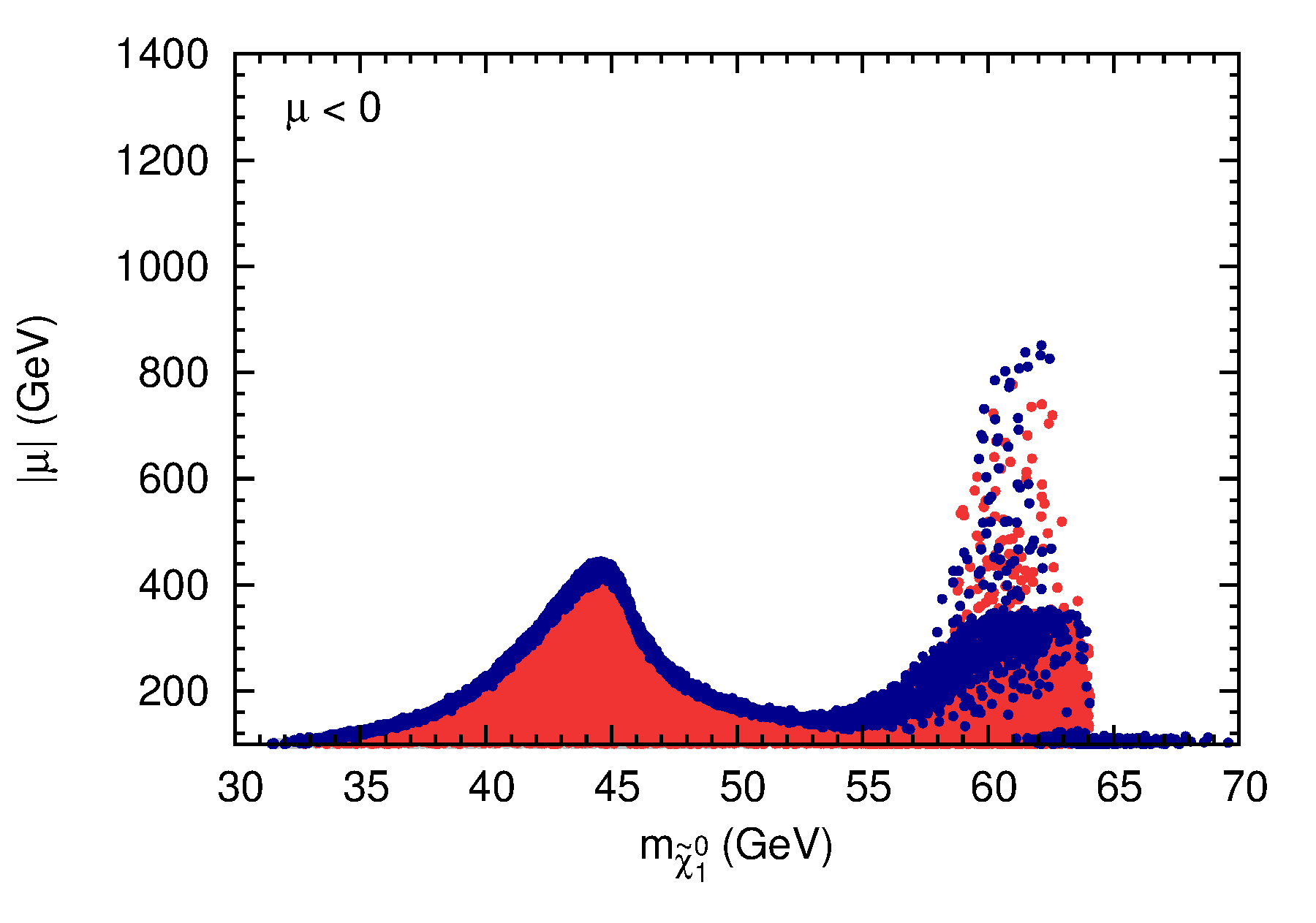}
 \caption{Results of the parameter scan defined in Eqs.~(\ref{eq:scanparameters}-\ref{eq:fixedparameters}) in the $m_{\nt_1}-|\mu|$ plane for $\mu>0$ (left panel), $\mu<0$ (right panel). Red points satisfy the relic density upper bound of Eq.~(\ref{eq:planck}) and all other constraints discussed in the text. Blue points in addition satisfy the lower bound. Gray points are excluded by one of the constraints listed in the text.}
 \label{fig:resonances} 
\end{figure}

The constraints we impose on our parameter space are presented in the
following.
\begin{itemize}
 \item {\bf DM relic density.} We assume a standard thermal history of the universe and 
 take this conservative range from Ref.~\cite{Ade:2013zuv}:
\begin{equation}
0.10 \le \Omega_{\rm DM}h^2 \le 0.13.
\label{eq:planck}
\end{equation}
\item {\bf Direct SUSY searches at LEP.} 
The 95\% CL bound on the lightest chargino mass is
\begin{equation}
\label{eq:LEPlimits}
m_{\ch_1} \ge 94~\text{GeV}.
\end{equation} 
Searches for $\nt_1 \nt_{2,3}$ associated production at LEP,
followed by the decay $\nt_{2,3} \to \nt_1 Z^{(*)}$, set a
constraint for $m_{\nt_1} + m_{\nt_{2,3}} \ge \sqrt{s} = 208$
GeV. This conservatively reads:
\begin{equation}
\sum_{k=2,3} \sigma (e^+ e^- \to \nt_1 \nt_k) \times {\rm BR}(\nt_k \to \nt_1 Z^{(*)}) < 100~{\rm fb}.  
\label{eq:opal}
\end{equation}
We estimated the production cross sections at LEP using the
leading order formulae reported in Refs.~\cite{Ellis:1983er,Bartl:1986hp}.
\item {\bf $Z$ invisible width.} 
As discussed above, the relic density constraint require
sizeable $\nt_1 \nt_1 Z$ and $\nt_1 \nt_1 h$ couplings. As a
consequence, the invisible decays $Z\to \nt_1\nt_1$ and $h\to
\nt_1\nt_1$ can occur at relevant rates if kinematically
allowed. 
The decay width of the $Z$ boson into a neutralino pair
is given by~\cite{Barbieri:1987hb}:
\begin{equation}
  \label{eq:Zinv}
    \Gamma (Z \rightarrow \widetilde{\chi}^{0}_{1} \widetilde{\chi}^{0}_{1}) =
  \frac{G_{F} m_Z^3}{12 \sqrt{2} \pi}  \left( 1 - \frac{4
      m_{\widetilde{\chi}^{0}_{1}}^{2}}{m_{Z}^{2}} \right)^{\frac{3}{2}} \left|
    N_{13}^{2} - N_{14}^{2} \right|^{2} \,,
\end{equation}
This has to be compared to the LEP bound on the new physics
contribution to $\Gamma(Z \to {\rm
invisible})$~\cite{ALEPH:2005ab}:
\begin{align}
 \Delta \Gamma_Z^{\rm inv} < 3  ~{\rm MeV}\quad(95\%~{\rm CL}).
\end{align}
\item{\bf Higgs mass and rates.}
Applying the tools HiggsBounds \cite{Bechtle:2008jh,Bechtle:2011sb,Bechtle:2013wla} 
and HiggsSignals \cite{Bechtle:2013xfa} 
we calculate a $\chi^2$ measure for the predictions of the model and the measured Higgs rates and mass. 
We ensure an agreement between the predicted light Higgs mass and
production rates and the current experimental measurements at the 95\% CL 
requiring a p-value below 0.002.

%
\begin{figure}[t]
\centering
 \includegraphics[width=0.48\textwidth]{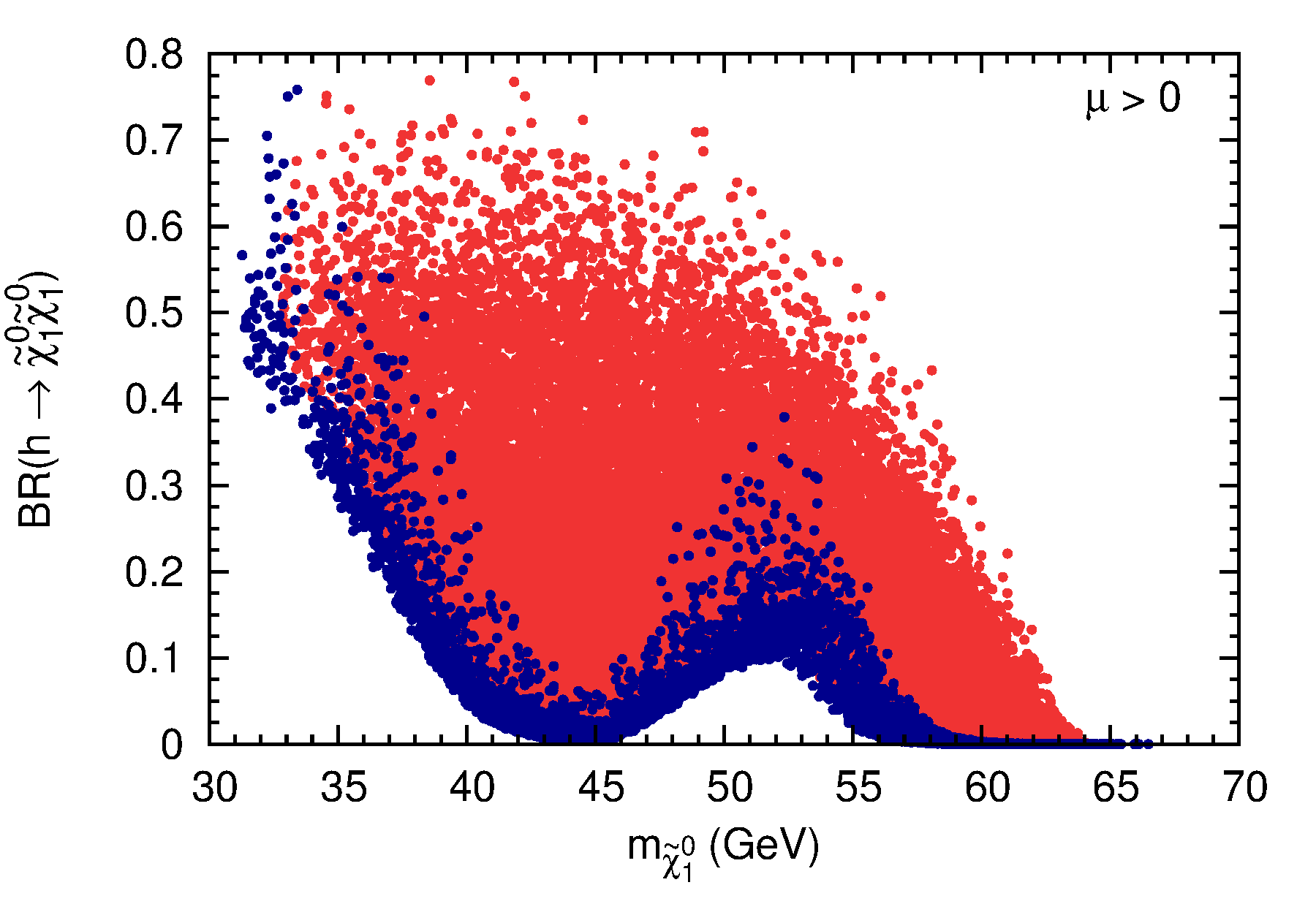}
 \includegraphics[width=0.48\textwidth]{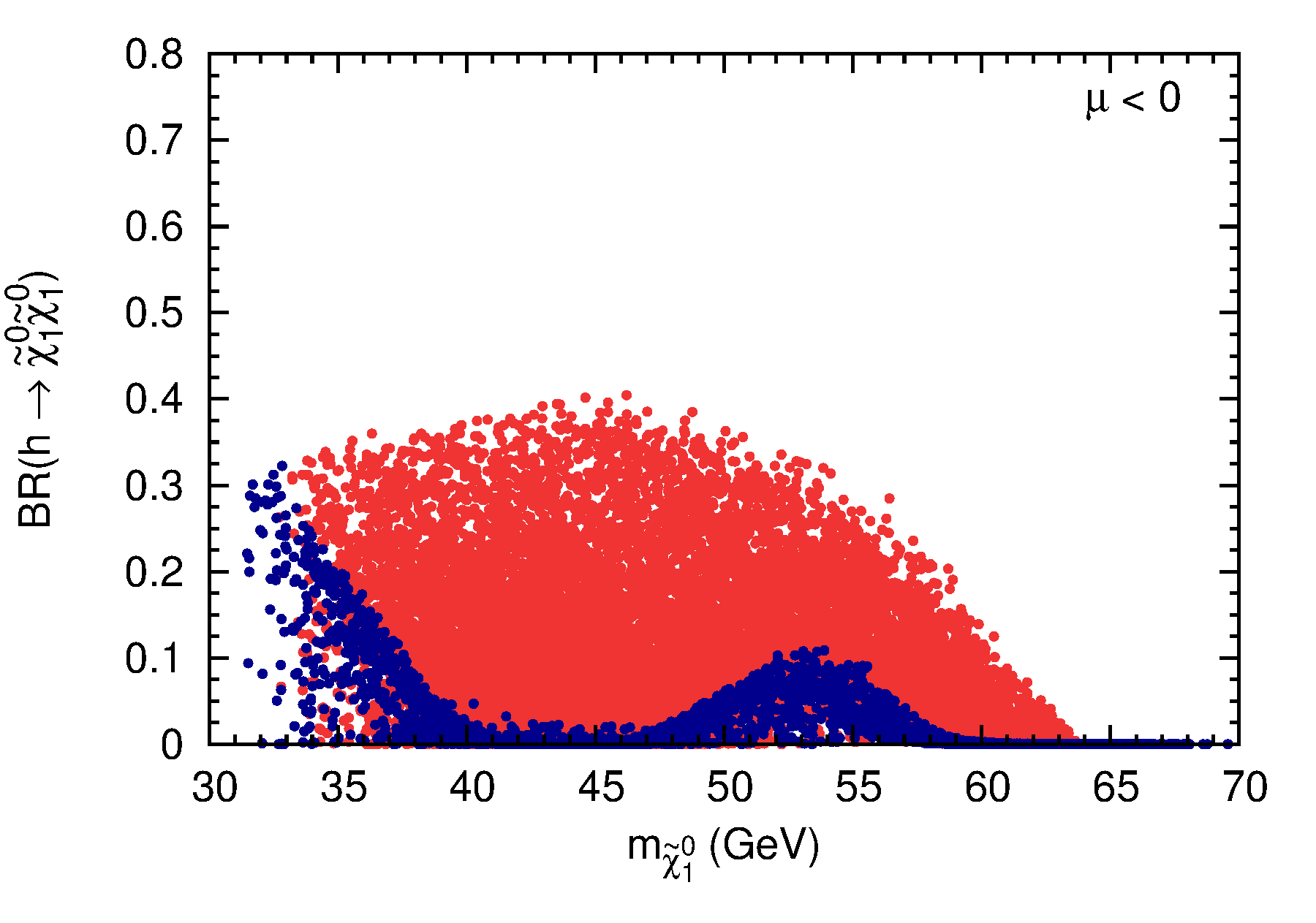}
  \caption{Predictions for the invisible branching ratio of the Higgs 
BR$(h \rightarrow \nt_1 \nt_1)$ in the paramter scan defined in Eqs.~(\ref{eq:scanparameters}-\ref{eq:fixedparameters}) for $\mu>0$ (left panel), $\mu<0$ (right panel). Red points satisfy the relic density upper bound of Eq.~(\ref{eq:planck}) and all other constraints discussed in the text. Blue points in addition satisfy the lower bound.}
 \label{fig:BRinv} 
\end{figure}
%
\item {\bf Invisible Higgs decays.}
The light Higgs decay width into $\nt_1\nt_1$ is given 
by~\cite{Griest:1987qv}:
\begin{equation}
\Gamma (h \rightarrow \nt_1 \nt_1)= \frac{\sqrt{2} G_F m_W^2 m_h}{ \pi}~
\left(  1- \frac{4 m_{\nt_1 }^2}{m_h^2}   \right)^{\frac{3}{2}} 
\big\vert  C^h_{11}   \big\vert^2 \;,
\end{equation}
where from Eq.~(\ref{eq:couplh}) one finds in the decoupling regime $m_A \gg m_h$:
\begin{equation}
 C^h_{11} = \frac{1}{2}\big( N_{12} -\tan \theta_W \; N_{11}   \big)
\big( \sin\beta\; N_{14} -\cos\beta \; N_{13}   \big)\;.
\end{equation}
Since a sizeable $\Gamma (h \to {\rm invisible})$ would reduce
by the same amount the branching fractions of all visible
channels, it can be constrained by fits to the observed Higgs
decay rates. In this work we adopt the limit reported in \cite{Bechtle:2014ewa}:
\begin{align}
 {\rm BR}(h \to {\rm invisible}) \lesssim 26 \%\quad(95\%~{\rm CL}). 
\label{eq:inv}
\end{align}
\end{itemize}

Possible further constraints from electroweak precision
observables or the flavour sector can be circumvented adjusting the 
parameters in Eq.~(\ref{eq:fixedparameters}).

In Fig.~\ref{fig:resonances} we show the results of the parameter scan in the plane of 
$m_{\nt_1}$ against the Higgsino mass parameter
$\mu$ for both signs of $\mu$. 
The red points only fulfill the
upper bound of Eq.~(\ref{eq:planck}), while the blue ones fulfill the
lower bound too. Thus the blue points correspond to models where $\nt_1$ can
account for 100\% of the observed Dark Matter.
Points excluded by any of the constraints explained above but the relic density constraint
are shown in grey and -- marginalizing over all other parameters -- they affect 
the parameter space only for $\mu > 0$ at small values of $\mu$. 

As argued already in the
previous section, if $m_{\nt_1}$ is slightly away from the resonances,
Eq.~(\ref{eq:planck}) tightly constrains $\mu$. On the other hand, Higgsinos
can be as heavy as $\approx 450$ GeV close to the $Z$-pole and as heavy as
$\approx  1200\, (900)$ GeV close to the $h$ resonance for $\mu > 0$ ($\mu < 0$).
The width and shape of the Higgs resonance is determined by the possible spread in the Higgs mass. 
Clearly, the parameter region very close to the $h$ resonance is difficult to cover 
entirely at the LHC.

For illustration in Fig.~\ref{fig:BRinv}, we explicitly show the invisible branching ratio of the Higgs
 BR$(h \rightarrow \nt_1 \nt_1)$ for both signs of $\mu$ obtained in our parameter scan. 
All constraints but the one from the invisible width of the Higgs
itself are applied and the color-coding is as in Fig.~\ref{fig:resonances}.  
As we can see, Eq.~(\ref{eq:inv}) excludes points for $m_{\nt_1}\lesssim 35$
GeV if $\mu>0$. For $\mu < 0$ no such limit can be obtained. 
In fact,  as one can see from Eq.~(\ref{eq:Higgsino}), a partial cancellation in the $\nt_1\nt_1 h$ vertex
 decreases the coupling if there is a relative sign between $\mu$ and $M_1$. 
 This is also the reason why smaller values of $|\mu|$ are required
  close to the $h$ resonance for $\mu < 0$,  see Fig.~\ref{fig:resonances}. 
 We want to note that
for the considered parameter space, regions excluded from the invisible width of the Higgs 
encompass exclusions from all other constraints considered here.

\subsection{Direct and indirect DM searches}
%
\begin{figure}[t]
\centering
  \includegraphics[width=0.48\textwidth]{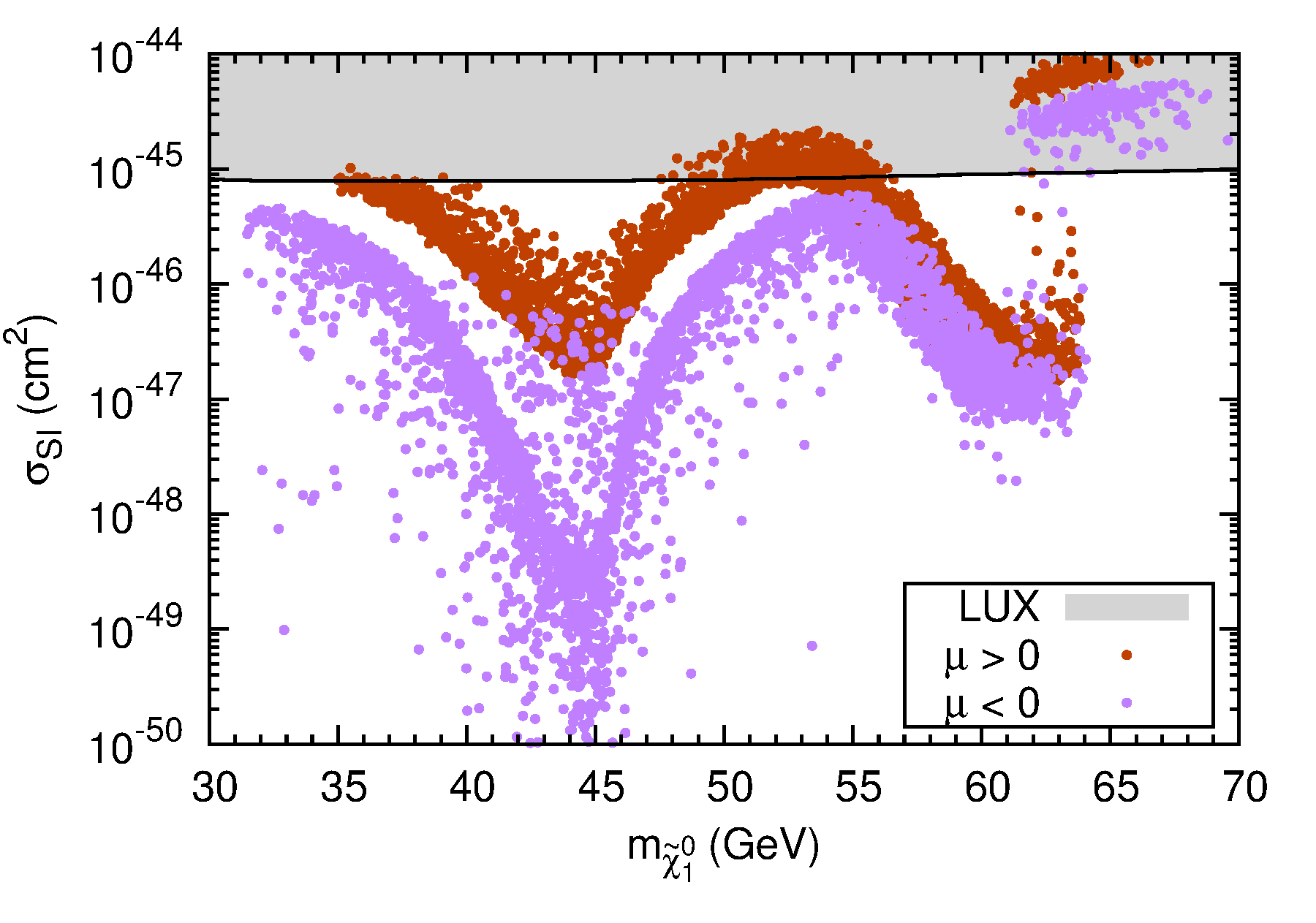}
 \includegraphics[width=0.48\textwidth]{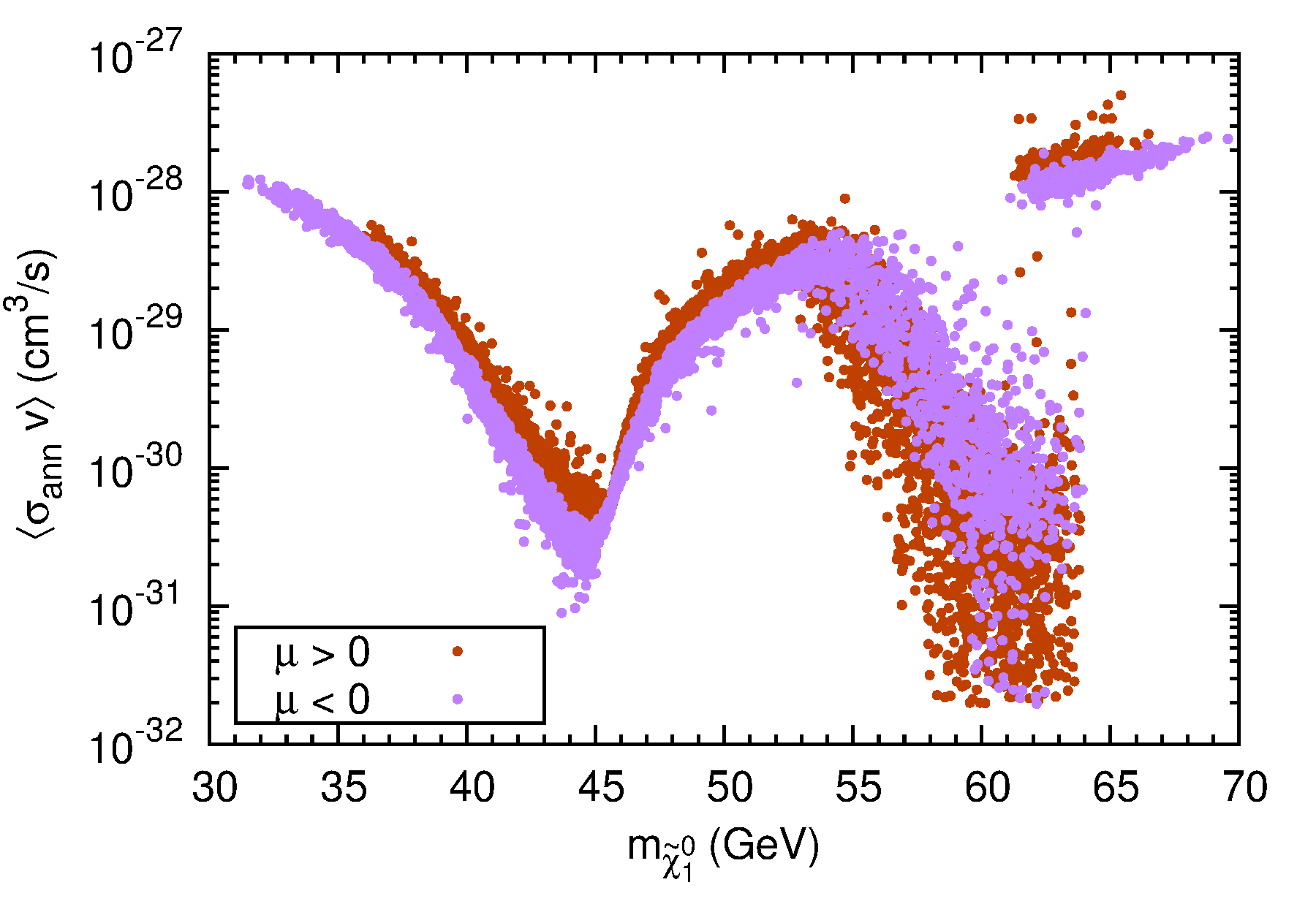}
  \caption{Predictions for the spin-independent DM-nucleon scattering cross section
$\sigma_{\rm SI}$ (left panel) and the present thermally-averaged annihilation cross section
$\langle\sigma_{\rm ann}v\rangle$ (right panel) in the paramter scan defined in Eqs.~(\ref{eq:scanparameters}-\ref{eq:fixedparameters}).
All points satisfy the upper and lower  bound of Eq.~(\ref{eq:planck}) and all other 
constraints discussed in section~\ref{sec:parameterspace}. The gray shaded area 
in the left plot is excluded from direct DM searches with LUX.}
 \label{fig:dir-ind} 
\end{figure}

As sketched in the introduction, direct and indirect DM 
searches can loose their sensitivity in the vicinity of the 
resonant annihilation regimes, Eq.~(\ref{eq:resonances}).
We quantify this behaviour in Fig.~\ref{fig:dir-ind},
where the spin-independent DM-nucleon scattering cross section
$\sigma_{\rm SI}$ (left panel) and the present
thermally-averaged annihilation cross section
$\langle\sigma_{\rm ann}v\rangle$ (right panel) are plotted as a
function of the lightest neutralino mass for the points of our parameter scan 
defined in Eqs.~(\ref{eq:scanparameters}-\ref{eq:fixedparameters}).
All shown points account for the observed DM abundance, i.e. they
satisfy the upper and lower bound of Eq.~(\ref{eq:planck}) besides 
all other constraints discussed in section~\ref{sec:parameterspace}.
Red (purple) points correspond to $\mu>0$ ($\mu<0$).

In the left panel, we show as a reference the current limit set
by the direct search experiment LUX \cite{Akerib:2013tjd},
which for the considered mass range is almost independent 
of the neutralino mass at 
$\sigma_{\rm SI} \lesssim 8\times10^{-46}~\mathrm{cm}^2$.
Close to the resonances the predicted $\sigma_{\rm SI}$ is suppressed 
by several orders of magnitude and tests of such scenarios 
 even in future direct DM search experiments seems to be 
very challenging. 
The neutralino elastic scattering with nuclei is
mediated by the exchange of CP-even Higgs states
or squarks (which we assume to be decoupled). 
Thus, the shown suppression originates from
small Higgs-Higgsino-bino couplings, as given in Eq.~(\ref{eq:couplh}),
close to the resonances (due to large $\mu$ as required by the relic abundance).
 Larger values of this coupling, i.e. a smaller $\mu$ parameter would
reduce the neutralino density $\Omega_{\nt}$ below the observed
value, which would require extra DM components and, more importantly for
us, anyway would reduce the sensitivity of direct detection by a
factor $\Omega_{\nt}/\Omega^{\rm obs}_{\rm DM}$.
However, we have to keep in mind that the
theoretical prediction for $\sigma_{\rm SI} $ suffers from large uncertainties:
variations of light quark masses and hadronic form factors, as
well as heavier values of $m_H\approx m_A$ (here we took $m_A=4$
TeV) can further reduce the predicted $\sigma_{\rm SI}$ by a
factor of few. On the other hand, lighter heavy Higgs states, i.e.~smaller 
values of $m_{A}$ could in principle increase the spin-independent
cross section without altering much the relic density prediction.
Therefore, we refrain from setting any conservative
constraints on our parameter space from direct detection.

Similarly to the discussion above, we observe in the right 
panel of Fig.~\ref{fig:dir-ind} that the predicted 
$\langle\sigma_{\rm ann}v\rangle$ is well below the 
sensitivity of indirect detection experiments -- which
are currently at the level of $10^{-26}$ cm$^3$/s \cite{PhysRevLett.107.241302} -- and
further drops in the vicinity of the resonances. 
The reason why the present $\langle\sigma_{\rm ann}v\rangle$ is 
much lower than the value required by a thermal WIMP at the
freeze-out can be understood by the following:
at high temperatures the annihilation is dominated by resonant $p$-wave
contributions which become irrelevant as the temperature drops.
In the present universe, the annihilation occurs through a
$Z$-mediated $s$-wave amplitude. The corresponding cross section is 
suppressed by a factor $m_f^2/m^2_Z$. Furthermore,
as above, close to the resonances a small Higgsino
component in $\nt_1$ further suppresses the $\nt_1\nt_1 Z$ coupling.
Again an additional possible contribution to $\langle\sigma_{\rm ann}v\rangle$ 
is expected to be provided by an $s$-channel exchange of a CP-odd Higgs $A$.
We checked that even for masses at the border of
the present LHC exclusion, e.g.~$m_A\simeq 500$ GeV for
$\tan\beta=20$ \cite{Khachatryan:2014wca}, $\langle\sigma_{\rm ann}v\rangle$
can not increase by more than one order of magnitude with
respect to the values shown in Fig.~\ref{fig:dir-ind}.

\section{LHC phenomenology}
\label{sec:pheno}

The spectrum we consider solely involves the neutralino/chargino
sector of the MSSM. 
As discussed above, the relic density
constraint translates into an upper bound on the Higgsino mass
parameter $\mu$, while the wino mass parameter $M_2$ does hardly
play a role satisfying those bounds. Thus, the minimal
particle content are just the mostly bino-like neutralino LSP
and the Higgsino states: two heavier neutralinos and the lightest
charginos.
Later we will demonstrate that additional light winos 
just increase the LHC sensitivity. Hence, taking
$M_2$ to be large is a conservative assumption and will be
assumed if not otherwise stated. 
All other SUSY particles are assumed to be decoupled.

For the described spectrum possible tests at the LHC rely on 
electroweak Drell-Yan production of the Higgsino-like 
states:\footnote{Monojet searches for direct production of a pair of 
neutralino LSPs in association with a jet can in principle also test the 
given spectrum but will only become sensitive in the future
~\cite{Han:2013usa,Low:2014cba}. }
\begin{equation}
 pp \,\to\, \nt_{k}\nt_{l},\quad pp \,\to\, \chp_{1}\chm_{1},\quad pp \,\to\, \ch_{1}\nt_{k},\qquad(k,l=2,3).
 \label{eq:production}
\end{equation}
The produced charginos can only decay into the LSP and (on- or off-shell) $W$ bosons:
\begin{equation}
\ch_{1}\,\to\, W^{\pm(*)} \nt_1,
\end{equation}
whereas the neutralinos have two competing decay modes, $Z$ or $h$:
\begin{equation}
\label{ew:neudecaymodes}
\nt_{2,3}\,\to\, Z^{(*)} \nt_1,\quad \nt_{2,3}\,\to\, h^{(*)}  \nt_1
\end{equation}
with relevance depending on the model parameters as discussed in the following.

\begin{figure}[t]
\centering
   \includegraphics[width=0.96\textwidth]{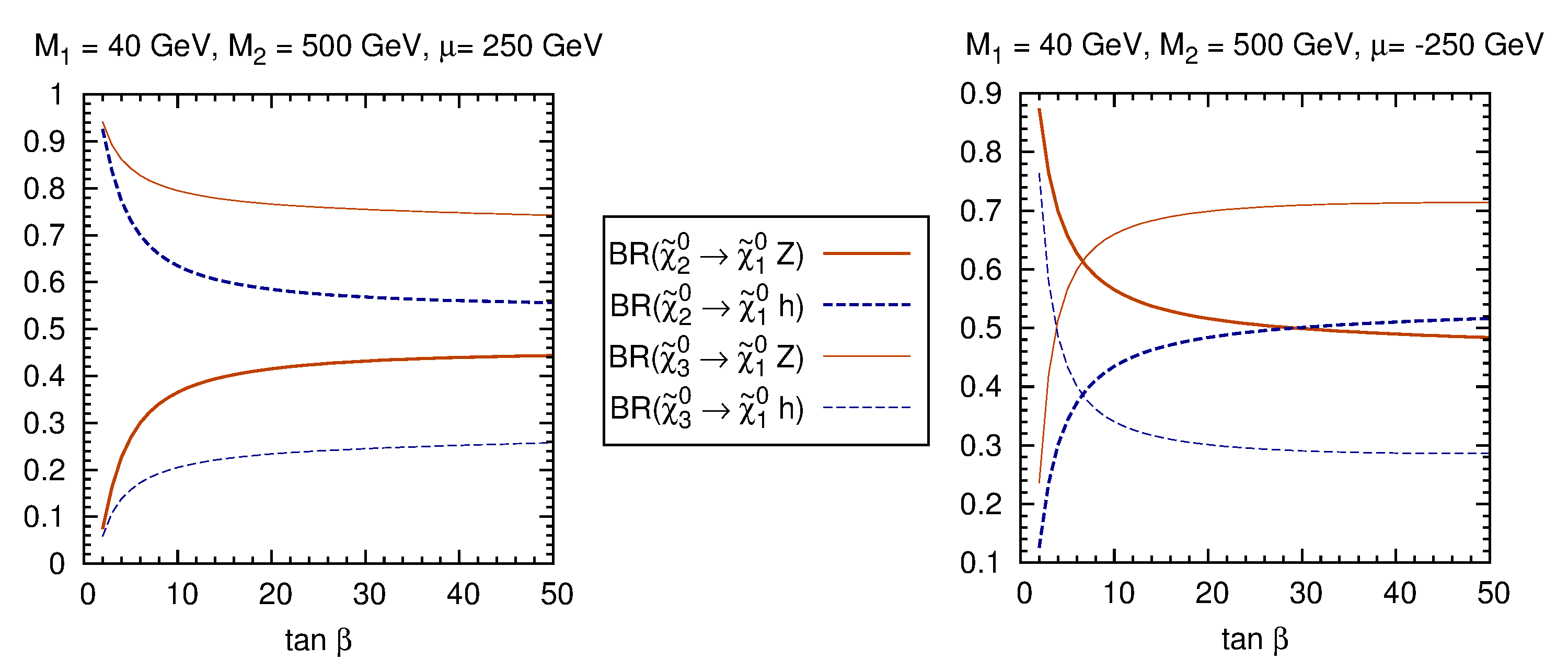}
 \caption{Branching ratios for the decay $\nt_{2,3} \to \nt_1 Z/h$ as function of  $\tan\beta$.
Parameters are chosen to be $M_1=40\gev, M_2=500\gev$ and $\mu=\pm250\gev$ for the
left/right plot.
}
 \label{fig:tb-BR}
\end{figure}

The most relevant searches for neutralino/chargino production
performed by the LHC collaborations are based on leptonic decays
of the gauge bosons, i.e.~on events with multiple leptons plus
missing transverse momentum. For the parameter space we consider 
by far the highest sensitivity is reached in the $WZ$-channel \cite{Martin:2014qra}
(from associated neutralino-chargino production, $\ch_{1}\nt_{2,3}$) 
with three reconstructed leptons in the final state \cite{Aad:2014nua,Khachatryan:2014qwa}. 
In this channel the search performed by ATLAS \cite{Aad:2014nua} 
sets the most stringent limits.
Searches for the Higgs decay have been performed in the
$Wh$-channel with $h\to b\bar b$ \cite{Khachatryan:2014qwa}, see also \cite{Baer:2012ts,Han:2013kza}.
The complementary $h\to \tau^+ \tau^-$ channel might yield a
similar sensitivity~\cite{Papaefstathiou:2014oja}. However, the overall sensitivity in the  
$Wh$-channel is considerably weaker compared to the $WZ$-channel. 
Still, it is in order to investigate in detail the rates of the competing decay 
modes shown in Eq.~(\ref{ew:neudecaymodes}).
The decay rates of $\nt_{2,3}\,\to\, Z \nt_1$ are controlled 
by the couplings defined in Eq.~(\ref{eq:couplZ}). 
Using the approximate expressions for the neutralino mixing
in the Higgsino-like $\nt_{2,3}$ limit $M_2\gg|\mu|$, as reported in 
the appendix Eq.~(\ref{eq:N-1th}), we find:
\begin{align}
 \mu>0:\quad &
 O^{ZL}_{21} \simeq \frac{m_{Z} s_{W}^{} }{2\sqrt{2} \mu }(s_{\beta} - c_{\beta})
\left( 1 + \frac{M_{1}}{\mu} \right),&
O^{ZL}_{31} \simeq \frac{m_{Z} s_{W}^{} }{2\sqrt{2} \mu }(s_{\beta} + c_{\beta})
\left( 1 + \frac{M_{1}}{\mu} \right); \\
\mu<0:\quad &
 O^{ZL}_{21} \simeq -\frac{m_{Z} s_{W}^{} }{2\sqrt{2} \mu }(s_{\beta} + c_{\beta})
\left( 1 + \frac{M_{1}}{\mu} \right), &
O^{ZL}_{31} \simeq \frac{m_{Z} s_{W}^{} }{2\sqrt{2} \mu }(c_{\beta} - s_{\beta})
\left( 1 + \frac{M_{1}}{\mu} \right),
\end{align}
Analogous expressions for the coupling $\nt_{2,3} \nt_1 h$ can
be obtained from Eqs.~(\ref{eq:couplh}, \ref{eq:N-1th}):
\begin{align}
 \mu>0:\quad &
 C^{h}_{21} \simeq -\frac{1}{2\sqrt{2}}(c_\beta +s_\beta) ,&
C^{h}_{31}  \simeq -\frac{1}{2\sqrt{2}}(c_\beta -s_\beta); \\
\mu<0:\quad &
 C^{h}_{21}\simeq -\frac{1}{2\sqrt{2}}(c_\beta -s_\beta), &
 C^{h}_{31}\simeq -\frac{1}{2\sqrt{2}}(c_\beta +s_\beta).
\end{align}
From these expressions
we expect that for $\mu>0$ the branching ratio
BR($\nt_{2}\,\to\, Z \nt_1$) decreases for small values of
$\tan\beta$, and vanishes in the limit $\tan\beta\to1$. Whereas
the branching ratio BR($\nt_{3}\,\to\, Z \nt_1$) is maximized in
the low $\tan\beta$ regime.
This behaviour is depicted in the left panel of
Fig.~\ref{fig:tb-BR}, where for illustration we choose
$M_1=40\gev, M_2=500\gev$ and $\mu=250\gev$.
The behaviour described above is reversed for $\mu<0$, as
shown in the right panel of Fig.~\ref{fig:tb-BR}.
%
Clearly, the $WZ$ channel is expected to suffer a loss of
sensitivity in the low (large) $\tan\beta$ regime for $\mu>0$
($\mu<0$). However, as the behavior of the two Higgsino-like
neutralinos $\nt_{2,3}$ is antipodal\footnote{Also the production cross sections
depend on $\tan\beta$  via the $\nt_{2,3}\ch_1Z$ couplings. For fixed physical masses 
the dependence is very mild and again antipodal to the corresponding 
dependence in the branching ratios.} and their mass splitting is in general small, 
the $\tan\beta$ dependence in the total sensitivity of the $WZ$ channel is moderate. 
In our numerical analysis in section \ref{sec:limits} we consider the two
example values $\tan\beta=5,40$.
Furthermore, as the summed contribution only mildly depend on $\tan\beta$
even for small values of $\tan\beta$ the $WZ$-channel is expected to remain 
more sensitive than the $Wh$-channel.

\begin{figure}[t]
\centering
  \includegraphics[width=0.65\textwidth]{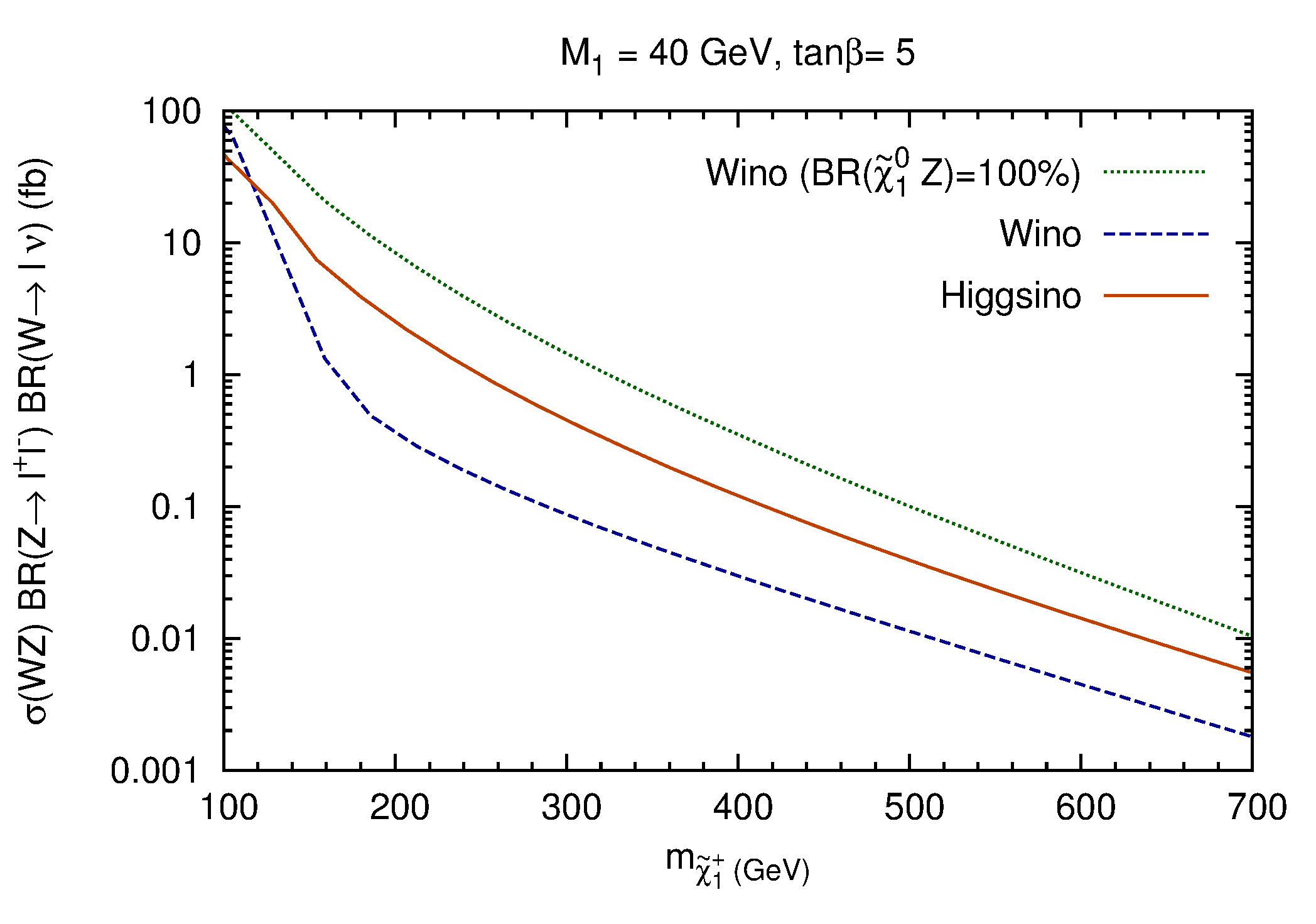}
 \caption{Comparison among the simplified wino model with BR($\nt_{2}\,\to\, Z \nt_1$)=100\%, a realistic
wino model with $M_2 \ll \mu$ and an Higgsino model with $\mu \ll M_2$. Shown is the summed 
neutralino-chargino production cross section times branching ratios into $W(\to \ell^{\pm}\nu)Z(\to \ell^+\ell^-)$ as defined in Eq.~(\ref{eq:prodBR}).}
 \label{fig:comparison}
\end{figure}
Searching for neutralino and chargino production in the three leptons
plus missing energy final state performed by ATLAS \cite{Aad:2014nua}
 the strongest available limits for the $WZ$ channel have been obtained.
These limits have been interpreted in the $M_2-\mu$ plane of the pMSSM
for fixed values of $M_1$ and in terms of constraints on the mass of purely wino-like
charginos and neutralinos with BR$(\nt_2\to \nt_1 Z)=100~\%$.
Clearly, the latter is only a simplified model as for a pure wino-like
$\nt_2$ state the coupling $\nt_1\nt_2 Z$ vanishes.
The corresponding limits for a realistic scenario with Higgsino-like
neutralinos and charginos might be much weaker due to changes 
in the cross section and possible competing decay modes as discussed above.
Therefore, in section \ref{sec:limits} we will reinterpret those 
limits for Higgsino-like neutralinos in a detailed analysis including
detector effects. 
Here, we already want to anticipate those results qualitatively.
To this end we compare the production cross section times branching ratio
for the $WZ$ channel defined as 
\begin{align}
\sigma_{3\ell+\missingET} =
\sum_{\substack{k=1,2\\l=2,3,4}}\sigma(\ch_k \nt_{l}) \, \BR(\ch_{l} \to W^\pm \nt_1)\, \BR( \nt_{l} \to Z \nt_1) 
\, \BR(W^\pm\to \ell^\pm \nu)  \, \BR(Z\to \ell^+\ell^-),
\label{eq:prodBR}
\end{align}
for the cases of (i) the simplified model,
(ii) realistic wino-like $\nt_{2}$ case,
and (iii) Higgsino-like $\nt_{2,3}$ case.
The corresponding estimated rates are shown in
Fig.~\ref{fig:comparison} as a function of the mass of $\ch_1$, 
where (leading order) production cross sections are calculated
with \prospino \cite{Beenakker:1999xh} and BRs with \susyhit \cite{Djouadi:2006bz}. 
In the Higgsino-like $\nt_{2,3}$ case, 
we set $M_2=1\tev$ and, vice versa, we set 
$\mu=1$ TeV in the wino-like $\nt_{2}$ cases. 
For all scenarios we additionally
set for illustration $M_1=40\gev$ and $\tan\beta=5$.
Clearly, for realistic values of BR($\nt_{2}\,\to\, Z \nt_1$) compared 
to the simplified model the sensitivity is strongly reduced 
in the wino-like $\nt_{2}$ case. 
Resulting rates are here also considerably smaller than in the 
Higgsino-like $\nt_{2,3}$ case -- despite the fact that 
the production cross section for wino states is typically larger 
than the one for Higgsinos.
In \cite{Aad:2014nua} ATLAS obtains (approximately) the limit 
$m_{\ch_1}\gtrsim 350$ GeV for $m_{\nt_1}\lesssim 100$ GeV under
the assumption of BR($\nt_{2}\,\to\, Z \nt_1$)=100\%.
Comparing such a limit in Fig.~\ref{fig:comparison} to the
Higgsino-like $\nt_{2,3}$ case we expect an exclusion on $m_{\ch_1}\approx
m_{\nt_2}\approx m_{\nt_3}$ weaker by about 100 GeV. Still, this
represents a non-negligible constraint on the neutralino DM parameter space.

%
\begin{figure}[t]
\centering
  \includegraphics[width=0.65\textwidth]{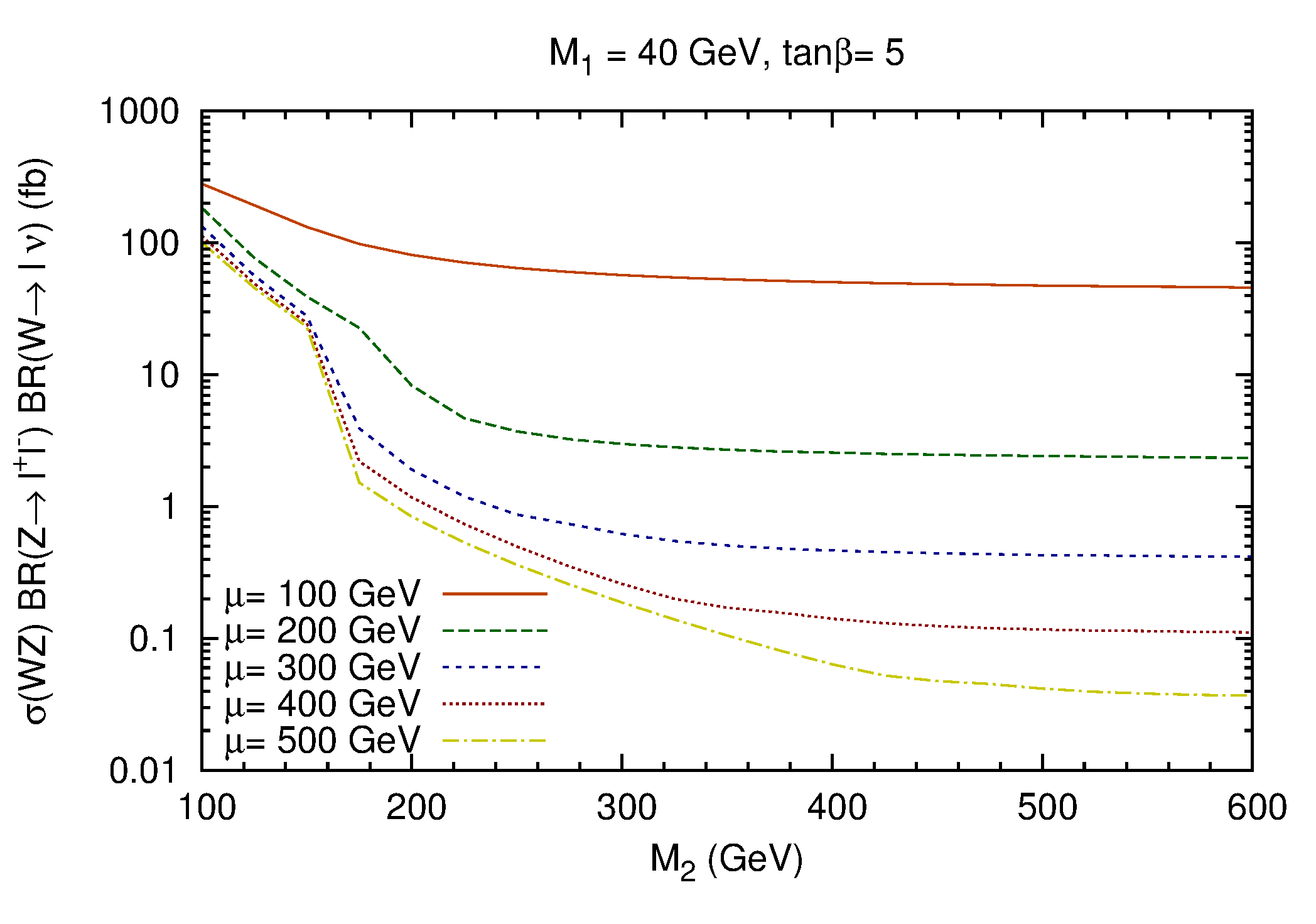}
 \caption{Impact of $M_2$ on the summed 
neutralino-chargino production cross section times branching ratios into $W(\to \ell^{\pm}\nu)Z(\to \ell^+\ell^-)$ as defined in Eq.~(\ref{eq:prodBR}) for different values of $\mu$.}
\label{fig:M2-sigmaBR}
 \end{figure}
%
Finally, let us turn to a short discussion of the possible
impact of the wino mass $M_2$ on the signal rates in the $WZ$
channel.
For $M_2 \lesssim |\mu|$, the full set of neutralinos and charginos
contributes to the production cross section, providing additional modes to those 
shown in Eq.~(\ref{eq:production}). 
Hence it is natural to expect an increase in sensitivity. 
This is confirmed and quantified in Fig.~\ref{fig:M2-sigmaBR}, where we plot
again cross section times branching ratio, as defined in Eq.~(\ref{eq:prodBR}),
now as a function of $M_2$ for different choices of $\mu$. 
In Fig.~\ref{fig:M2-sigmaBR} we see that the number of expected leptonic events 
generically increases for low values of $M_2$, whereas it becomes approximately 
flat for  $M_2>\mu$. As a consequence, fixing $M_2$ at some value larger than $\mu$
can be regarded as a conservative choice. We are going to adopt this choice in the numerical 
simulation of the next section.

Let us note here that further MSSM parameters, besides
those of the neutralino/chargino sector, cf.~Eq.~(\ref{eq:parameters}), 
have in general little impact on the searches based on electroweak production 
of Higgsinos and their decay. In particular, the
production cross sections have no dependence on any squark masses,
in contrast to the wino case, for which $t$-channel squark
exchange decouples only very slowly and can be relevant even 
for very heavy squarks~\cite{Papaefstathiou:2014oja}. 
In contrast to the wino case the Higgsinos have only
small couplings with first and second generation quarks and
squarks rendering such contributions negligible.

\section{LHC limits}
\label{sec:limits}

\begin{figure}[t]
\centering
 \includegraphics[width=0.65\textwidth]{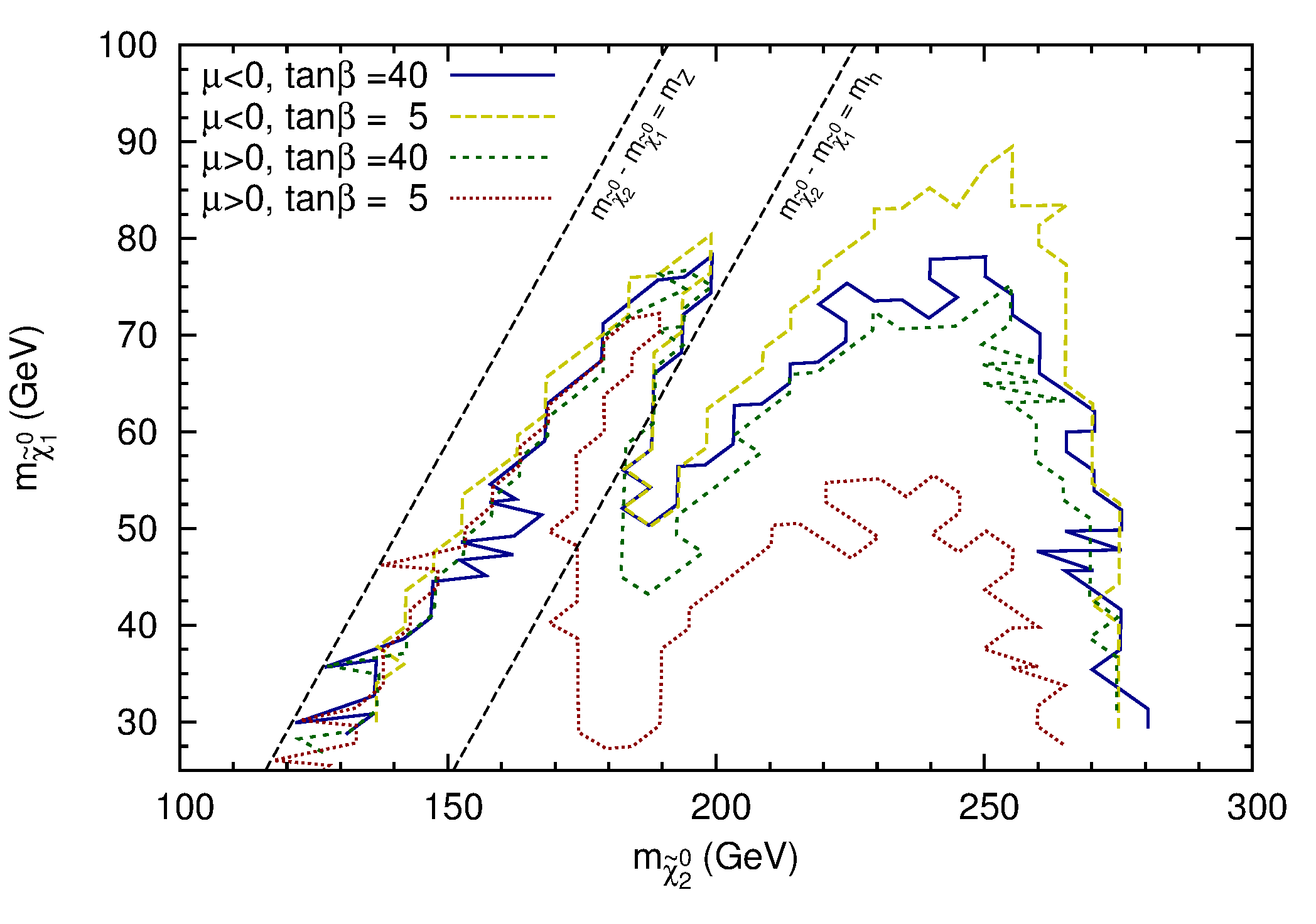}
 \caption{Reinterpreted ATLAS limit \cite{Aad:2014nua} for the Higgsino-like $\nt_{2,3}$ case,
 displayed in the $m_{\nt_2}$-$m_{\nt_1}$) plane for different values of $\tan\beta$, ${\rm sgn}(\mu)$.
 See the text for details.}
\label{fig:results_first}
\end{figure}

As discussed in the last section, the scenario under
consideration can best be searched for at the LHC in the $WZ$ channel, 
where both CMS and ATLAS have performed
different searches using the full dataset available at 8~TeV
\cite{Aad:2014nua,Aad:2014vma,Khachatryan:2014qwa}. The most
stringent limit available is deduced from the three-leptons plus
missing energy search performed by ATLAS \cite{Aad:2014nua}. In
the relevant signal region three leptons have to be identified,
where two of them have to be of the same flavour and of
different sign (SF-OS). The resulting event sample is further
divided into 16 bins with different invariant mass cuts for the
SF-OS pair, different cuts on the transverse mass $m_T$ and/or
different cuts on the transverse missing energy $\missingET$.
Final event numbers are found to be in good agreement with
Standard Model predictions. Interpreting the resulting limits in
a pure wino scenario\footnote{In this scenario the sfermions are
decoupled at $m_{\tilde f}=5$~TeV as listed on HepData \cite{HepData}.} with
BR($\nt_{2}\,\to\, Z \nt_1)=100\%$ (as explained above) ATLAS
sets bounds up to $m_{\ch_1}=m_{\nt_2} \gtrsim 350~$GeV for a
massless neutralino. Furthermore, the ATLAS collaboration
interprets the search in the $M_2$-$\mu$-plane of a pMSSM
scenario with decoupled sfermions, a bino of $M_1=50~$GeV
and $\tan\beta=10$. Here, for $M_2 \gg \mu$ a limit of $\mu
\gtrsim 230~$GeV is derived. We want to reinterpret this limit
in the light neutralino scenario discussed above, where we vary
both $M_1$ and $\tan\beta$ (besides $\mu$ and $M_2$).

In our Monte Carlo study we use \herwig \cite{Bellm:2013lba} for
event simulation and rescale the obtained LO rates 
with NLO K-factors obtained from \prospino \cite{Beenakker:1999xh}.
Furthermore, we use the powerful \checkmate \cite{Drees:2013wra}
framework for detector simulation (where a tuned version of
\delphes \cite{deFavereau:2013fsa} is used internally), analysis
and statistical evaluation. First, we carefully verified that
the three-leptons plus missing energy analysis implemented in
\checkmate yields limits for the pure-wino scenario and the
pMSSM scenario which are in good agreement with the ones
published by ATLAS. Second, we evaluate those limits in the
benchmark scenarios motivated in section~\ref{sec:pheno}, 
i.e.~we translate the ATLAS limits into the $M_1-\mu$ plane for
$\tan\beta=5,40$ and $\mu \gtrless 0$. As discussed in section
\ref{sec:pheno}, for $\mu>0$, $\tan\beta=5$ gives a conservative limit while
$\tan\beta=40$ gives a limit in the plateau shown in Fig.
\ref{fig:tb-BR}. On the contrary,  for $\mu<0$, the large $\tan\beta$
case corresponds to a conservative scenario.
As also discussed in section \ref{sec:pheno}
decoupling $M_2$ yields a conservative bound and thus we set
$M_2 =1$~TeV.

Resulting limits are shown in Fig. \ref{fig:results_first}
projected on the plane of the physical masses $m_{\nt_1}$ vs.
$m_{\nt_2}$. For a lightest neutralino of mass $m_{\nt_1}=35~$GeV,
$\tan\beta=5$ and $\mu>0$ we find a limit of $m_{\nt_2} \lesssim
120,~m_{\nt_2} \gtrsim 260~$GeV. The lower limit is a consequence of
the kinematic boundary between on- and off-shell decays at
$m_{\nt_2} -m_{\nt_1}=m_Z$. In the regime of purely off-shell decays of
the $\nt_2$ various decay modes compete and the ATLAS limit
vanishes. Furthermore, as discussed in \cite{Bharucha:2013epa}, in
this regime branching ratios can still strongly depend on
the scale and details of the ``decoupled'' sfermions, thus
conservative exclusion limits are difficult to deduce. For
$\tan\beta=40$ and $\mu<0$ the upper limits extends up to $m_{\nt_2}
\gtrsim 280~$GeV. At the same time in this case larger values of
$m_{\nt_1}$ can be excluded. For all scenarios studied exclusion
limits drop sharply at the kinematical threshold $m_{\nt_2}-m_{\nt_1} =
m_h$. In this small corner of the parameter space limits from
$Wh$ searches might become relevant \cite{Baer:2012ts,Han:2013kza,Papaefstathiou:2014oja}.

\begin{figure}[t]
\centering
 \includegraphics[width=0.48\textwidth]{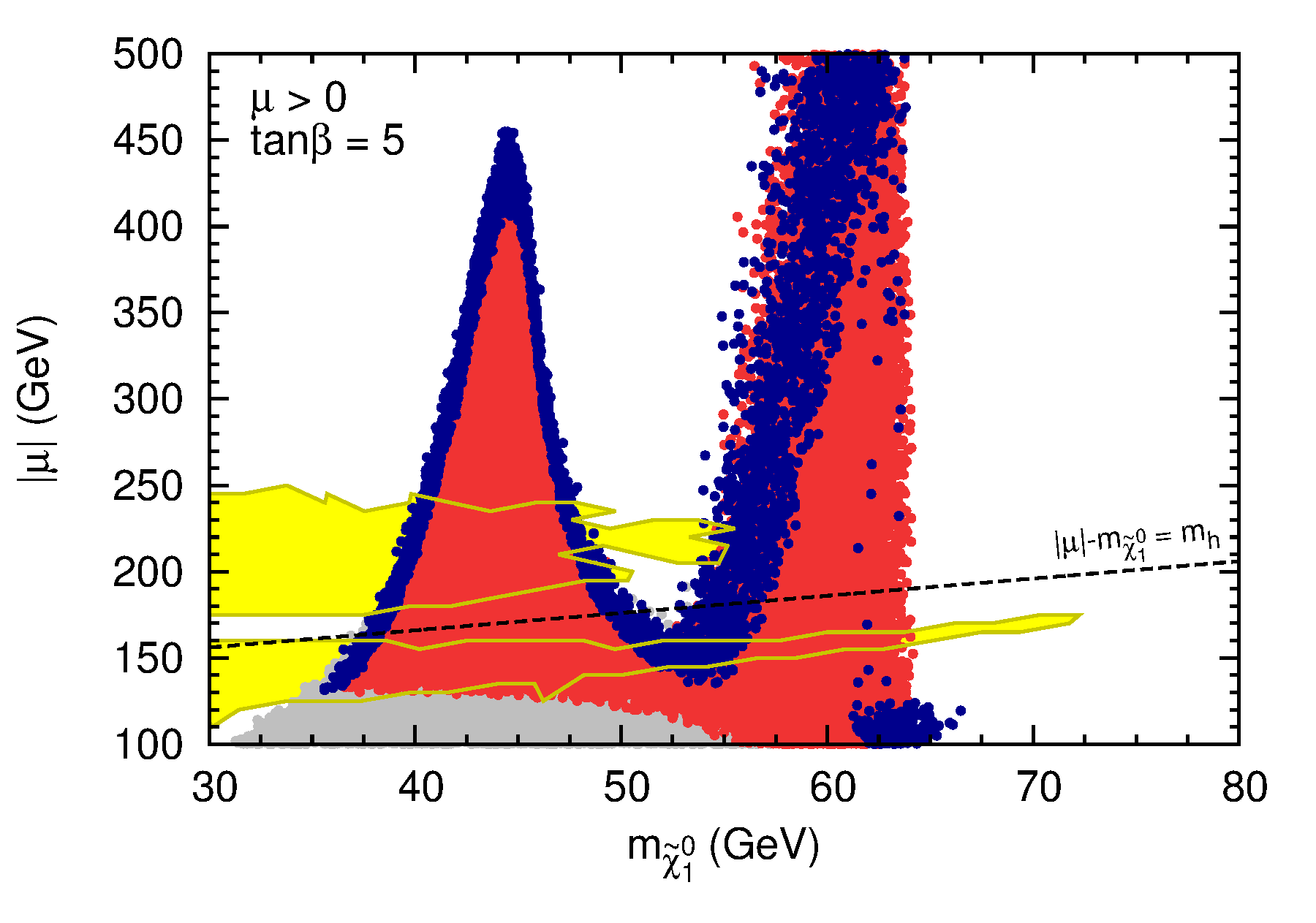}
 \includegraphics[width=0.48\textwidth]{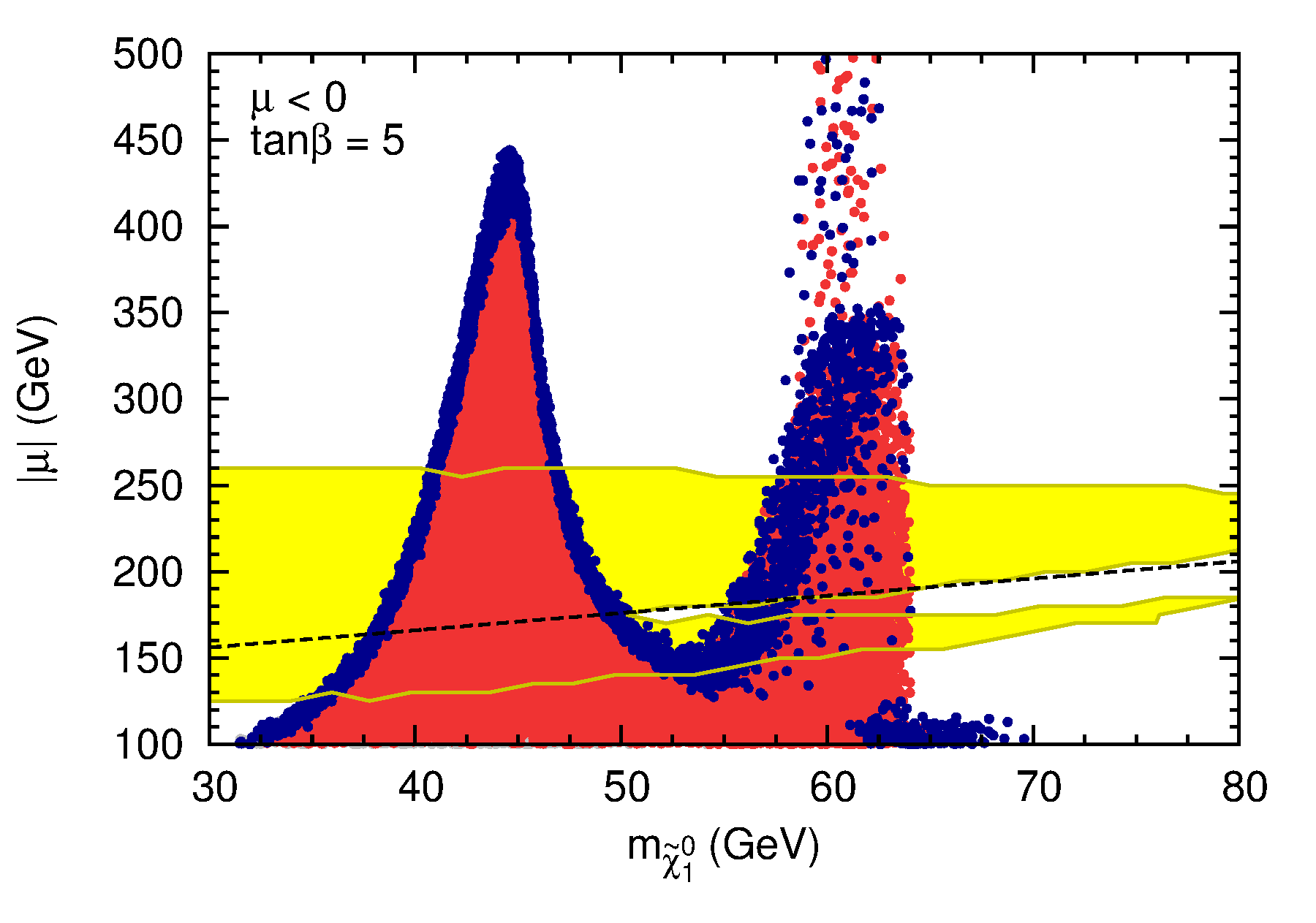}
 \includegraphics[width=0.48\textwidth]{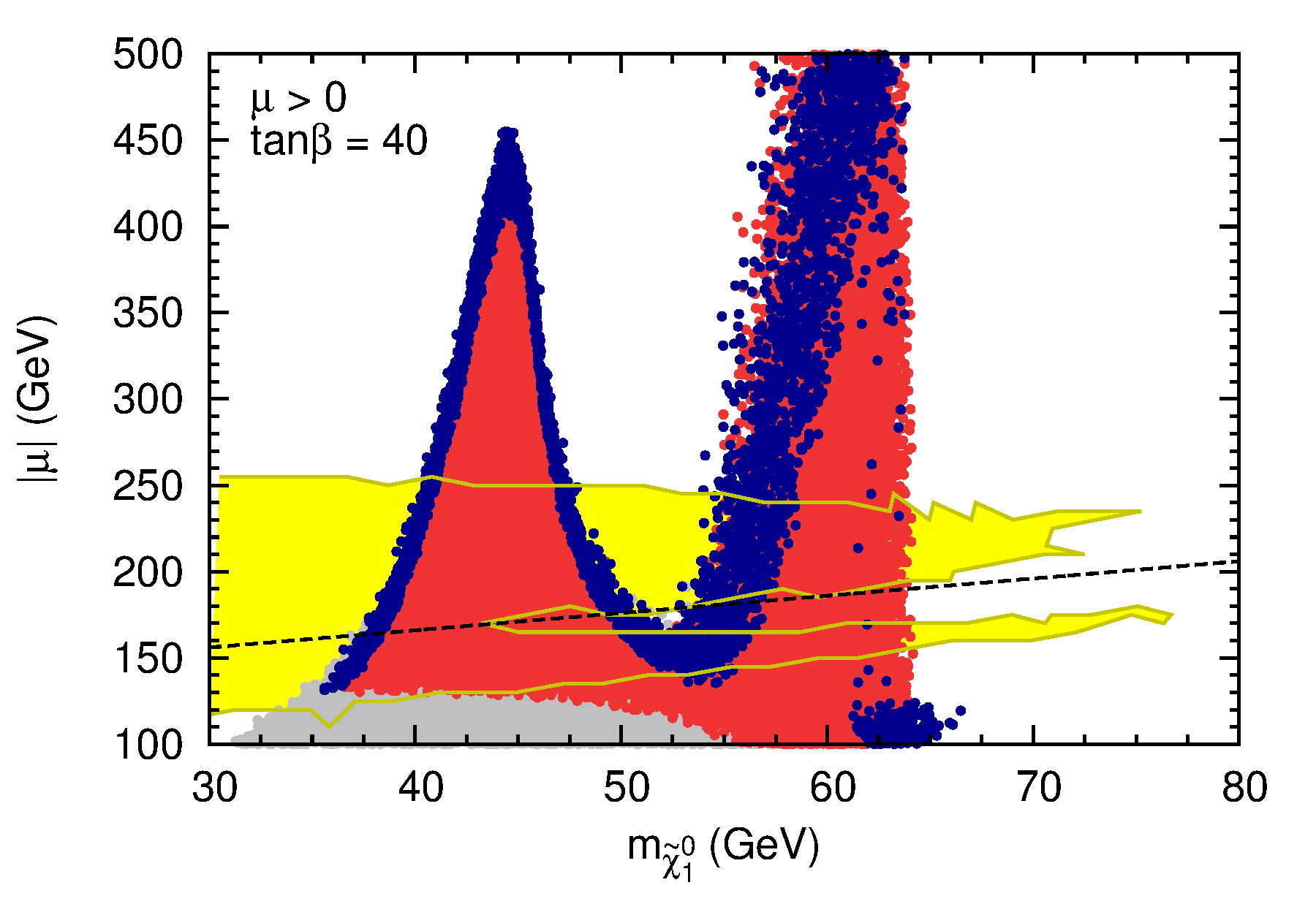}
 \includegraphics[width=0.48\textwidth]{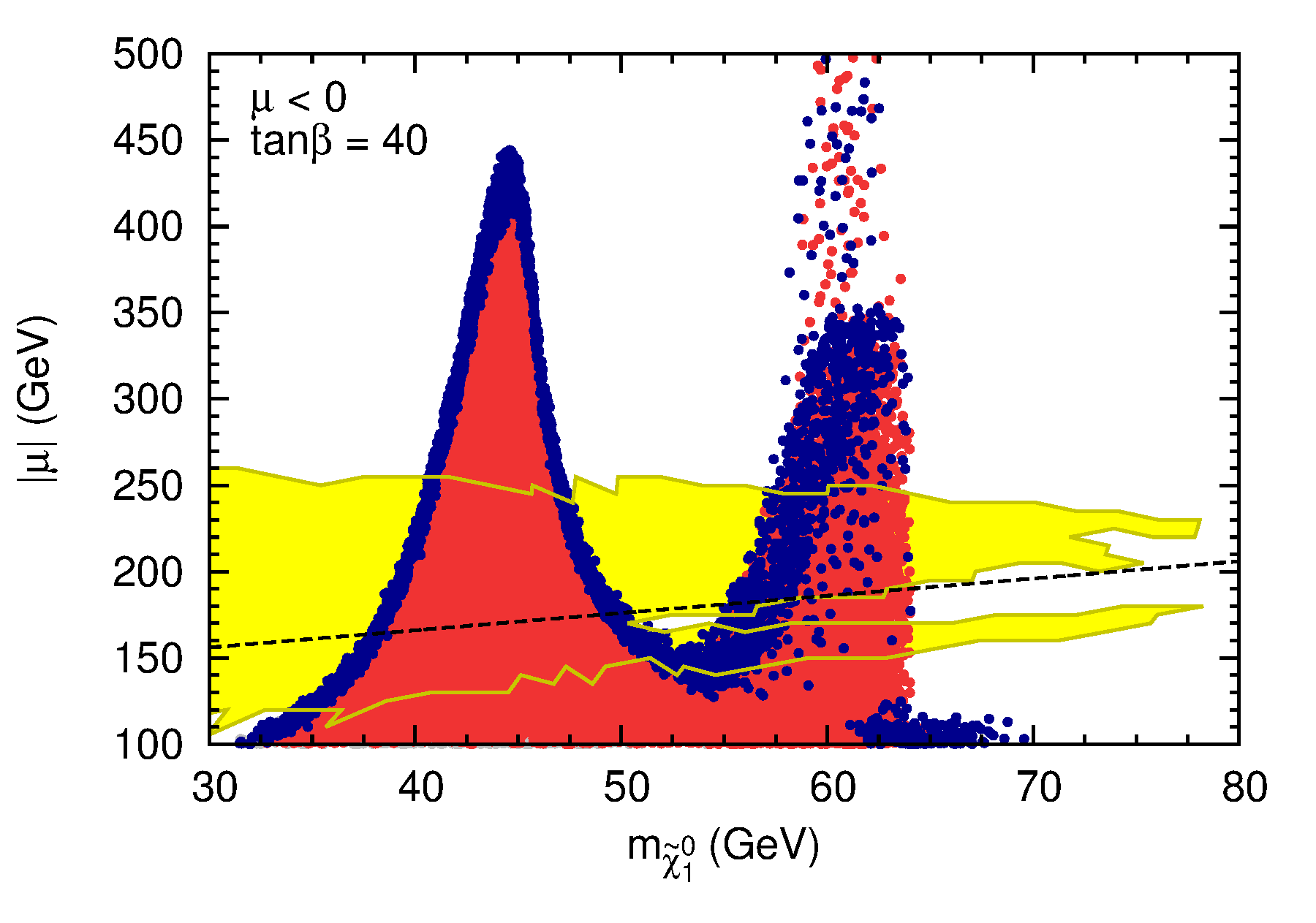}
 \caption{Reinterpreted ATLAS limit,
 displayed in the $m_{\nt_1}-|\mu|$ plane of Fig.~\ref{fig:resonances} for different values of $\tan\beta$, ${\rm sgn}(\mu)$.
  Color code as in  Fig.~\ref{fig:resonances}.
 The yellow regions are excluded by ATLAS three leptons plus missing energy search \cite{Aad:2014nua}.
}
\label{fig:results_final}
\end{figure}

Finally we present the limits of our reinterpretation in the
phenomenologically interesting $\nt_1$ vs.~$\vert \mu \vert$
plane, i.e.~in the plane where constraints from the thermal
relic abundance were discussed in Fig.~\ref{fig:resonances}.
Results are shown in the upper/lower panel of Fig.~\ref{fig:results_final} for $\tan\beta =5/40$; on the left and
right for $\mu>0$ and $\mu<0$ respectively. Regions excluded by
LHC searches are shaded in yellow and as in Fig.~\ref{fig:resonances} regions yielding the correct relic
abundance just from neutralino DM are shown in blue, while
regions where the abundance of the $\nt_1$ could contribute to
the overall DM abundance are shown in red. 
For all choices of $\tan\beta$ and $\text{sgn}(\mu)$ large parts of the
$Z$-resonance regions are excluded. More precisely, at the
$Z$-resonance we find $\mu\gtrsim 250$~GeV, apart from a small
strip around ($|\mu|\approx m_{\nt_2})-m_{\nt_1} = m_h$ for $\mu > 0$
and in general for very small values of $\mu$ (below the
$\nt_2\to Z\nt_1$ threshold). However, at least for $\mu>0$,
here the limit from $ {\rm BR}(h \to {\rm invisible})$ shown in~Eq.~(\ref{eq:inv}) become relevant, excluding the points shown in grey.
 Combining
these limits for $\tan\beta=40$ we find
\begin{align}
m_{\nt_1} \gtrsim 40~\text{GeV} \quad [\mu>0,~\tan\beta=40]\, .
\end{align} 
In the more conservative scenario with $\tan\beta=5$ the bound
is somewhat weaker and we find
\begin{align}
m_{\nt_1} \gtrsim 37~\text{GeV}\quad [\mu>0,~\tan\beta=5] \, .
\end{align} 
For $\mu<0$ the constraint from the relic abundance combined with the LEP limit on charginos 
still yields the strongest bound as very small values of $\mu$
cannot be excluded. As we have already seen in section
\ref{sec:parameterspace} here,
\begin{align}
m_{\nt_1} \gtrsim 30~\text{GeV} \quad [\mu<0]\, .
\end{align}

Also the $h$-resonance region can partly be excluded already.
Precise limits can be read of Fig.~\ref{fig:results_final}.
Noteworthy, this region extents to very large
values of $\mu$, beyond the scope of even the high energy LHC.

\section{Conclusions}
\label{sec:conclusions}

In this work, we have studied light neutralino Dark Matter in the MSSM
within frameworks where all sfermions are heavy.
Interestingly, this feature of the spectrum is shared by both `natural
SUSY' and `mini split' scenarios.
Under the assumption that the neutralino is a standard thermal relic,
CMB measurements of the DM abundance translate
into specific requirements the spectrum must fulfill. Since, in our
case, sfermions play no role in the computation of
the DM relic density, these bounds must be satisfied by the
neutralino/chargino sector of the MSSM alone.
The generic requirement is that Higgsinos are relatively light, such
that the lightest neutralino can couple to $Z$ or $h$
through a non-negligible Higgsino component and thus efficiently annihilate.
This condition is strongly relaxed if the neutralino mass approaches
the conditions for a resonant enhancement
of the annihilation cross section: $m_{\nt_1}\simeq m_Z/2$ or
$m_{\nt_1}\simeq m_h/2$.
In such a case, Higgsinos can be as heavy as 450 GeV and 1.2 TeV respectively.
This parameter space, depicted in Fig.~\ref{fig:resonances}, can be
hardly tested by direct and indirect DM search experiments
because of a suppression of the relevant cross sections in
correspondence of the resonances, as shown in Fig.~\ref{fig:dir-ind}.
In contrast, LHC experiments have the potential to partly test our scenario
searching for production of Higgsino-like charginos and neutralinos,
followed by decays to $WZ$ and the LSP.
In fact, the remarkable sensitivity reached by the LHC experiments in
the search for purely electroweakly interacting new particles
allows us to directly test the electroweak sector of supersymmetric
models without the need of assumptions on the strongly-interacting
superpartners.

In section \ref{sec:limits}, we have presented the results of our
reinterpretation of an ATLAS three leptons plus missing energy search.
In Fig.~\ref{fig:results_final}, we have shown that LHC experiments
can set non-trivial constraints on the light neutralino parameter
space
already with the data collected at $\sqrt{s}=8$ TeV. In particular, we
have seen that the current limit only leaves uncovered
the case of a neutralino mass lying close to the resonances (at about
5 GeV or less), as well as the corners of the parameter space
corresponding to the kinematical thresholds
$|\mu|-m_{\nt_1}=m_Z,~m_h$, where the three leptons searches loose
sensitivity.
Combining with further channels, such as di-leptons plus missing
energies, as well as searches for $Wh$ events, could further reduce
the uncovered corners.

The exercise we have performed demonstrates once more that LHC
searches for electroweakly interacting SUSY particles
can be successfully interpreted as indirect searches for
supersymmetric Dark Matter at collider (especially in combination
with the relic density constraints), often resulting in more stringent limits
than those set by Dark Matter experiments themselves.

\section*{Acknowledgments}
\noindent We thank Jamie Tattersall for providing us with a private version of \checkmate. 
We would also like to thank Andreas Papaefstathiou for useful discussions and help with \herwig.
We also thank Ulrich Ellwanger and Guillaume Drieu La Rochelle for valuable discussions.
JML was supported by the European Commission through the ``LHCPhenoNet''
Initial Training Network PITN-GA-2010-264564.
The research of TO is supported by Grants-in-Aid for Scientific Research 
on Innovative Areas {\it Unification and Development of the Neutrino
Science Frontier} Number {\sf 2610 5503}.

\appendix

\section{Neutralino masses and mixing}
\label{app:massesmixings}

The neutralino mass term in the MSSM Lagrangian
\begin{align}
 \mathscr{L}
 =&
 -
 \frac{1}{2}
 (\overline{\psi^{c}})_{\alpha}
 (\mathcal{M}_{\widetilde{\chi}^{0}})_{\alpha \beta}
 \psi_{\beta},
\end{align}
is given with the following symmetric matrix,
\begin{align}
 (\mathcal{M}_{\widetilde{\chi}^{0}})_{\alpha \beta} 
 = 
 \begin{pmatrix}
 M_{1} & 0 & - m_{Z} s_{W}^{} c_{\beta} & m_{Z} s_{W}^{} s_{\beta} 
\\
 0 & M_{2} & m_{Z} c_{W}^{} c_{\beta} & -m_{Z} c_{W}^{} s_{\beta} 
\\
-m_{Z} s_{W}^{} c_{\beta} & m_{Z} c_{W}^{} c_{\beta} & 0 & -\mu
\\
m_{Z} s_{W}^{} s_{\beta} & - m_{Z} c_{W}^{} s_{\beta} & -\mu & 0 
\end{pmatrix},
\end{align}  
in the gauge-interaction basis 
$\psi_{\alpha} = (\widetilde{B}, \widetilde{W}^{0},
\widetilde{H}_{d}^{0}, \widetilde{H}_{u}^{0})$,
where $s_{W}^{} = \sin \theta_{W}$, 
$c_{W}^{} = \cos \theta_{W}$, $s_{\beta} = \sin \beta $,
and $c_{\beta} = \cos \beta$.
Although we diagonalize this mass matrix (with the radiative
corrections) numerically to obtain the mass eigenvalues 
$m_{\widetilde{\chi}^{0}_{i}}$ and the mass eigenstates 
$\widetilde{\chi}^{0}_{i} = N_{i \alpha} \psi_{\alpha}$ 
with the mixing matrix $N_{i \alpha}$, 
we derive analytic and approximated expressions 
to grasp the trend of numerical results. 

First, we separate $\mathcal{M}_{\widetilde{\chi}^{0}}$ into 
the zero-th order part $\mathcal{M}_{0}$ and the perturbation
$\delta \mathcal{M}$ as $\mathcal{M}_{\widetilde{\chi}^{0}} 
= \mathcal{M}_{0} + \delta \mathcal{M}$. 
They are defined as
\begin{align}
 (\mathcal{M}_{0})_{\alpha \beta}
 =&
 \begin{pmatrix}
  M_{1} & 0 & 0 & 0
  \\ 
  0 & M_{2} & 0 & 0 \\
  0 & 0 & 0 & -\mu \\
  0 & 0 & -\mu & 0 
 \end{pmatrix},
 \\
 (\delta \mathcal{M})_{\alpha \beta}
 =&
  \begin{pmatrix}
 0 & 0 & - m_{Z} s_{W}^{} c_{\beta} & m_{Z} s_{W}^{} s_{\beta} 
\\
 0 & 0 & m_{Z} c_{W}^{} c_{\beta} & -m_{Z} c_{W}^{} s_{\beta} 
\\
-m_{Z} s_{W}^{} c_{\beta} & m_{Z} c_{W}^{} c_{\beta} & 0 & 0
\\
m_{Z} s_{W}^{} s_{\beta} & - m_{Z} c_{W}^{} s_{\beta} & 0 & 0 
\end{pmatrix}.
\end{align}
The zero-th order part $\mathcal{M}_{0}$ can be diagonalized 
with the zero-th order mixing matrix $N_{i \alpha}^{(0)}$
\begin{align}
 N_{i \alpha}^{(0)}
 =
 \begin{pmatrix}
  1 & 0 & 0 & 0
  \\
  0 & 0 & - \frac{1}{\sqrt{2}} & \frac{1}{\sqrt{2}}
  \\
  0 & 0 & - \frac{1}{\sqrt{2}} & - \frac{1}{\sqrt{2}}
  \\
  0 & -1 & 0 & 0 
 \end{pmatrix},
\label{eq:N-0th}
\end{align}
and the mass eigenvalues at the zero-th order are given as\footnote{%
At the zero-th order, $\widetilde{\chi}^{0}_{2}$ and $\widetilde{\chi}^{0}_{3}$ 
are degenerate in physical mass.
The {\it second lightest} state and the {\it third lightest} state 
can be identified only after taking the radiative corrections 
(and the perturbation) into account.
In the case of $\mu>0$, we identify the state with a mass 
of $\mu$ as a would-be $\widetilde{\chi}^{0}_{2}$,
and that with $-\mu$ as a would-be $\widetilde{\chi}^{0}_{3}$. 
In the case of $\mu<0$,
the ordering becomes opposite, i.e., 
$m_{\widetilde{\chi}^{0}_{2}}^{(0)} = - \mu (>0)$
and 
$m_{\widetilde{\chi}^{0}_{3}}^{(0)} = \mu (<0)$.
Therefore, the mixing matrix for the $\mu<0$ case 
can be obtained by exchanging $N_{2\alpha}$ and $N_{3\alpha}$ 
in Eq.~\eqref{eq:N-1th}.
}
\begin{align}
 m_{\widetilde{\chi}^{0}_{1}}^{(0)} = M_{1},
\quad
 m_{\widetilde{\chi}^{0}_{2}}^{(0)} = \mu,
\quad
 m_{\widetilde{\chi}^{0}_{3}}^{(0)} = -\mu,
\quad
 m_{\widetilde{\chi}^{0}_{4}}^{(0)} = M_{2}.
\end{align}
Since we assume $M_{1} \ll |\mu| \ll M_{2}$ in our scenario,
here we arrange the ordering of the mass eigenstates as
described with Eq.~\eqref{eq:N-0th}.

Next, we take into account the effect from the perturbation part 
$\delta \mathcal{M}$.
This perturbation is valid, if the condition 
$(\delta \mathcal{M})_{ij} \ll 
\Bigl|m_{\widetilde{\chi}^{0}_{i}}^{(0)} 
- m_{\widetilde{\chi}^{0}_{j}}^{(0)}\Bigr|$
is fulfilled, where $(\delta \mathcal{M})_{ij}$ 
is the perturbation part in the zero-th order mass eigenbasis,
i.e., 
$(\delta \mathcal{M})_{ij} 
= N^{(0)}_{i \alpha} (\delta \mathcal{M})_{\alpha \beta} 
N^{(0) {\sf T}}_{\beta j}$.
In the $M_{2}$ decoupling limit,
the perturbation terms are proportional to $m_{Z}^{} s_{W}^{}$,
which is sufficiently smaller than the difference between 
two mass eigenvalues, which is typically $|\mu|$.
After including the first order perturbation, 
the neutralino mixings $N_{i \alpha}$
result in the followings: 
\begin{align}
\hspace{-0.5cm}
 N_{i \alpha}^{(0+1)}
 =
 \begin{pmatrix}
  1 & 0 & 
  \frac{m_{Z} s_{W}^{}}{\mu}
\left(
 s_{\beta} + c_{\beta} \frac{M_{1}}{\mu}
 \right) & 
 -
 \frac{m_{Z} s_{W}^{}}{\mu}
 \left(
 c_{\beta}
 +
 s_{\beta}
 \frac{M_{1}}{\mu}
 \right)
  \\
  \frac{m_{Z} s_{W}^{} (s_{\beta} + c_{\beta})}{\sqrt{2} \mu }
\left(
1 + \frac{M_{1}}{\mu}
\right) & 0 & - \frac{1}{\sqrt{2}} & \frac{1}{\sqrt{2}}
  \\
  \frac{m_{Z} s_{W}^{} (s_{\beta} - c_{\beta})}
{\sqrt{2} \mu }
\left(
1 - \frac{M_{1}}{\mu}
\right) & 0 & - \frac{1}{\sqrt{2}} & - \frac{1}{\sqrt{2}}
  \\
  0 & -1 & 0 & 0 
 \end{pmatrix},
\label{eq:N-1th}
\end{align}
Here, we expand the elements in powers of $M_{1}/\mu$ and leave 
the terms up to the first order.
They fit well with the numerical results evaluated by {\tt SuSpect}.
Although the mass eigenvalues do not get the correction 
at the first order of this perturbation, they are affected 
by the radiative corrections \cite{Fritzsche:2002bi}, 
which are larger than the perturbations.
For the analytic expressions of the neutralino masses and mixings 
in various cases, see e.g., Ref.~\cite{Bartl:1989ms}.

\bibliographystyle{apsrev}

\end{document}